\documentclass[10pt,preprint,a4paper]{aastex}
\usepackage{graphics,epsf}
\usepackage{amsmath}                
\usepackage{amsfonts}               
\usepackage{amssymb}                
\usepackage{epsfig}                 
\usepackage{subfigure}

\def \cm{~\rm{cm}}
\def \s{~\rm{s}}
\def \km{~\rm{km}}

\def \K{~\rm{K}}
\def \g{~\rm{g}}

\def \AU{~\rm{AU}}

\def \yr{~\rm{yr}}

\def \days{~\rm{day}}

\title{Bipolar rings from jet-inflated bubbles around evolved binary stars}

\author{Muhammad Akashi\altaffilmark{1} \& Noam Soker\altaffilmark{1}}

\altaffiltext{3}{Department of Physics, Technion -- Israel
Institute of Technology, Haifa 32000 Israel;
akashi@physics.technion.ac.il; soker@physics.technion.ac.il.}

\begin{document}

\begin{abstract}
We show that a fast wind that expands into a bipolar nebula composed of two opposite jet-inflated bubbles can form a pair of bipolar rings around giant stars.
Our model assumes three mass loss episodes: a spherical slow and dense shell, two opposite jets, and a spherical fast wind. We use the FLASH hydrodynamical code in three-dimensions to simulate the flow, and obtain the structure of the nebula.
We assume that the jets are launched from an accretion disk around a stellar companion to the giant star. The accretion disk is assumed to be formed when the primary giant star and the secondary star suffer a strong interaction accompanied by a rapid mass transfer process from the primary to the secondary star, mainly a main sequence star. Later in the evolution the primary star is assumed to shrink and blow a fast tenuous wind that interacts with the dense gas on the surface of the bipolar structure.
We assume that the dense mass loss episode before the jets are launched is spherically symmetric. Our results might be applicable to some planetary nebulae, and further emphasize the large variety of morphological features that can be formed by jets. But we could not reproduce some of the properties of the outer rings of SN~1987A. It seems that some objects, like SN~1987A, require a pre-jets mass loss episode with a mass concentration at mid-latitudes.
\end{abstract}


\section{INTRODUCTION}
\label{sec:intro}

Bipolar nebulae around evolved stars, such as planetary nebulae (PNe), are thought by many researchers to be shaped by jets that inflate bubbles inside a dense wind, (e.g,  \citealt{Moris1987, CorradiSchwarz1993, SahaiTrauger1998, GarciaSegura2005, Soker2007, GarciaDiaz2008, AkashiSoker2008a, Dennisetal2008, Dennisetal2009, Leeetal2009, Velazquezetal2011, Boffinetal2012, HuarteEspinosaetal2012, AkashiSoker2013, Balicketal2013, Thomasetal2013, Rieraetal2014, Tocknelletal2014, Velazquezetal2014, Zijlstra2015}). These bubbles can lead to other morphological features such as a dense equatorial outflow (\citealt{SokerRappaport2000, AkashiSoker2008b, Akashietal2015}). The rich variety of
outflow types, such as slow winds and fast winds that can be
blown, before, during, and after the jets-launching episodes, and
the rich variety of interacting binary \citep{Soker1998, DeMarco2015} and triple \citep{Soker2016} stellar systems ensure the rich variety of descendant nebular morphologies,
practically making almost every bipolar nebula `unique' \citep{Soker2002b}.

Some of these nebulae contain one pair of bipolar rings (double-rings) e.g., SN~1987A (e.g. \citealt{Burrowsetal1995, Panagiaetal1996, CrottsHeathcote2000, Maranetal2000, Sugermanetal2005}) and the PN He~113 \citep{Sahaietal2000}, while some others contain several rings in each lobe, e.g., the PNe MyCn~18 \citep{Sahaietal1999} and NGC~6881 \citep{KwokSu2005}, and the Red Rectangle post-AGB nebula \citep{Cohenetal2004}. In SN 1987A the two bipolar rings are referred to as the outer rings. The line connecting the centers of the rings is displaced from the exploding star. Several scenarios for the formation of the outer rings of SN~1987A have been proposed over the years (e.g., \citealt{Soker1999, Soker2002a, TanakaWashimi2002, Sugermanetal2005, MorrisPodsiadlowski2007, MorrisPodsiadlowski2009}). In the present study we aim at producing one pair of bipolar rings from a spherically symmetric fast wind that expands into bipolar jets-inflated bubbles.

Formation of rings by jets interacting with CSM were studied
before. \cite{Soker2002a} suggested the formation of double-ring
systems from jets interacting with thin dense shells. As a jet
interacts with the thin shell a high pressure region is formed,
that in turn accelerates the shells material sideways. This
process forms a ring. The dense shells are formed from an
impulsive mass-loss episode. According to this scenario, due to the binary interaction and its
orbital motion the double-ring system is displaced from the
symmetry axis of the main nebula, as observed in SN 1987A.
\cite{Soker2005} proposed a scenario where repeated jet-launching episodes from a mass-accreting stellar companion form the multiple double-ring system of the Red Rectangle. The jets expand into the dense AGB wind close to the binary system. \cite{Martinetal2015} argue that precessing jets are required to shape the Red Rectangle rather than wide jets. We note that the central star of the Red Rectangle is a post-AGB star (e.g., \citealt{VanWinckel2004}), and hence no phase of fast wind followed the formation of the rings. In the present study we include a fast wind, hence the presently studied scenario does not apply to the Red Rectangle.  \cite{Dennisetal2008} were aiming at several rings structure, and simulated the formation of rings by launching bullets instead of a continuous jet.  \cite{Dennisetal2009} further found that rings are formed in jet-inflated lobes when magnetic fields are included, but not without magnetic fields.

Although there are recent models for forming bipolar nebulae without jets (e.g., \citealt{Chenetal2016}), we attribute the formation of bipolar nebulae to jets.
As well, we assume that magnetic fields do not play a significant role in the shaping process far from the source of the winds and jets, and will present pure gas-dynamical simulations. The numerical set up is described in section \ref{sec:setup}, the results are presented in sections \ref{sec:rings} and \ref{sec:lobes}, and our short summary is in section \ref{sec:summary}.

\section{NUMERICAL SET UP}
\label{sec:setup}

Our simulations are performed by using version 4.0-beta of the
FLASH code \citep{Fryxell2000}. The FLASH code is an adaptive-mesh
refinement (AMR) modular code used for solving hydrodynamics or
magnetohydrodynamics problems. Here we use the unsplit PPM
(piecewise-parabolic method) solver of FLASH. We neither include
gravity, as velocities are much above the escape speed in the
region we simulate, nor radiative cooling,  as the interaction
region is optically thick. Instead of calculating radiative
cooling and radiative transfer, that are too complicated for the
flow geometry, we take low values of the adiabatic index, $\gamma =1.02$ and $\gamma =1.1$ to mimic radiative cooling via photon diffusion. Using lower values of $\gamma<5/3$ to mimic radiative cooling is reasonable when kinetic energy is
channelled to thermal energy, but not when thermal energy is
channelled to kinetic energy. For more details on the numerical
settings and the justifications for the parameters employed in
this study see \cite{AkashiSoker2013}.

We employ a full 3D AMR (8 levels; $2^{11}$ cells in each
direction) using a Cartesian grid $(x,y,z)$ with outflow boundary
conditions at all boundary surfaces. We take the $x-y$ plane with
$z=5\times10^{15} \cm$ to be in the equatorial plane of the PN,
and we simulate the whole space (the two sides of the equatorial
plane).

At $t=0$ we place a spherical dense shell in the region $R_{\rm
in} = 10^{14} \cm < r < 2\times 10^{14} \cm= R_{\rm out}$, and
with a density profile of $\rho_s = 4.74 \times 10^{-11}(r/10^{14}
\cm)^{-2} \g \cm^{-3}$, such that the total mass in the shell is
$0.3M_\odot$. The gas in the shell has an initial radial velocity
of $v_s = 10 \km \s^{-1}$. The shell corresponds to a mass loss
episode lasting for about $3 \yr$ and with a constant mass loss
rate of $\dot M_s \simeq 0.1 M_\odot \yr^{-1}$.  Such an event
can be classified as an intermediate luminosity optical transient (ILOT; \citealt{SokerKashi2012, AkashiSoker2013}, where more details can be found). The regions outside and inside the dense shell are filled with a lower density spherically-symmetric slow wind having a uniform radial
velocity of  $v_{\rm wind}=v_s= 10 \km \s^{-1}$. The slow wind
density at $t=0$ is taken to be $\rho(t=0) = {\dot M_{\rm
wind}}({4 \pi r^{2} v})^{-1}$, where $\dot M_{\rm wind}=10^{-5}
{\rm M_\odot \yr^{-1}}$.

The two opposite jets are lunched from the inner $10^{14} \cm $
region along the $z$-axis and  within a half opening angle of
$\alpha = 50^\circ$. By the term `jets' we refer also to wide
outflows, as we simulate here. More generally, we simulate
slow-massive-wide (SMW) outflows. Although the jets that are
simulated here are not observed (because the medium is optically
thick), such wide outflows are commonly observed to be blown from active
galactic nuclei (e.g., \citealt{Aravetal2013}). This, and the
success of wide outflows to explain lobes observed in cooling flow
clusters (e.g., \citealt{Sternbergetal2007}) motivate us to
consider wide outflows.

The jets' launching episode lasts for $5 \times 10^{6} \s = 58 \days$, coresponding to about one orbital period of a companion around a massive star of radius of $1 \AU$.
The jets' initial velocity is $v_{\rm jet}=700 \km \s^{-1}$, corresponding to the escape speed form a main sequence star, and the mass loss rate into the two jets together is
$\dot M_{\rm 2jets} = 0.13 M_\odot \yr^{-1}$. The high mass loss rate is assumed to occur as a companion enters or grazes the envelope of a giant star. 
The slow wind, dense shell, and the ejected jets start with a
temperature of $1000 \K$. The initial jets' temperature has no
influence on the results (as long as it is highly supersonic) because
the jets rapidly cool due to adiabatic expansion. For numerical
reasons a weak slow wind is injected in the sector
$\alpha<\theta<90^\circ$ (more numerical details are in \citealt{AkashiSoker2013}).

A new ingredient compared with our previous paper is a fast wind that starts to be blown at
$t=8 \times 10^{6} \s$. Its speed is $v_{\rm fast}=1000 \km \s^{-1}$
and the mass loss rate is $\dot M_{\rm fast} = 0.01 M_\odot \yr^{-1}$. A high mass loss rate is assumed to exist as the remnant of the two merged star is out of equilibrium for some time.
As we discuss below, in reality the fast wind is expected to last for a much longer time and its mass loss rate is expected to be lower. But for numerical reasons we simulate the fast wind to be a short episode just after the jets-launching phase.

\section{RAPID COOLING: THE FORMATION OF RINGS}
\label{sec:rings}

As described in section \ref{sec:setup} there are three consecutive flow components. (1) A slowly expanding spherical shell that is assumed to be blown by the primary giant star as a result of a short and strong binary interaction. (2) Two opposite jets that are assumed to be launched by a main sequence companion that accretes at a high rate from the primary envelope. In the results to be presented here this phase lasted in the period $t=0-58 \days$. (3) A fast wind blown by the primary star as a result of the merger process with the secondary star, or from the contraction of the star, as in the evolution to a blue giant of the progenitor of SN 1987A.
We present the interaction between these three components, but limit the presentation to emphasize the physics of the flow and the morphology.

For numerical reasons (saving resources) we launch the fast spherical wind starting at time
$t=8 \times 10^{6} \s = 0.25 \yr$. However, in reality the structure formed from the interaction of the jets with the shell can evolve for tens to hundreds of year before the fast wind has its full power.  For that, the present study cannot be compared quantitatively to observed objects. We limit the present study to the general morphology that can be formed by these three mass-loss episodes.

\begin{figure}
\subfigure[$t=0.25$~\yr]{\includegraphics[height=4in,width=4in,angle=0]{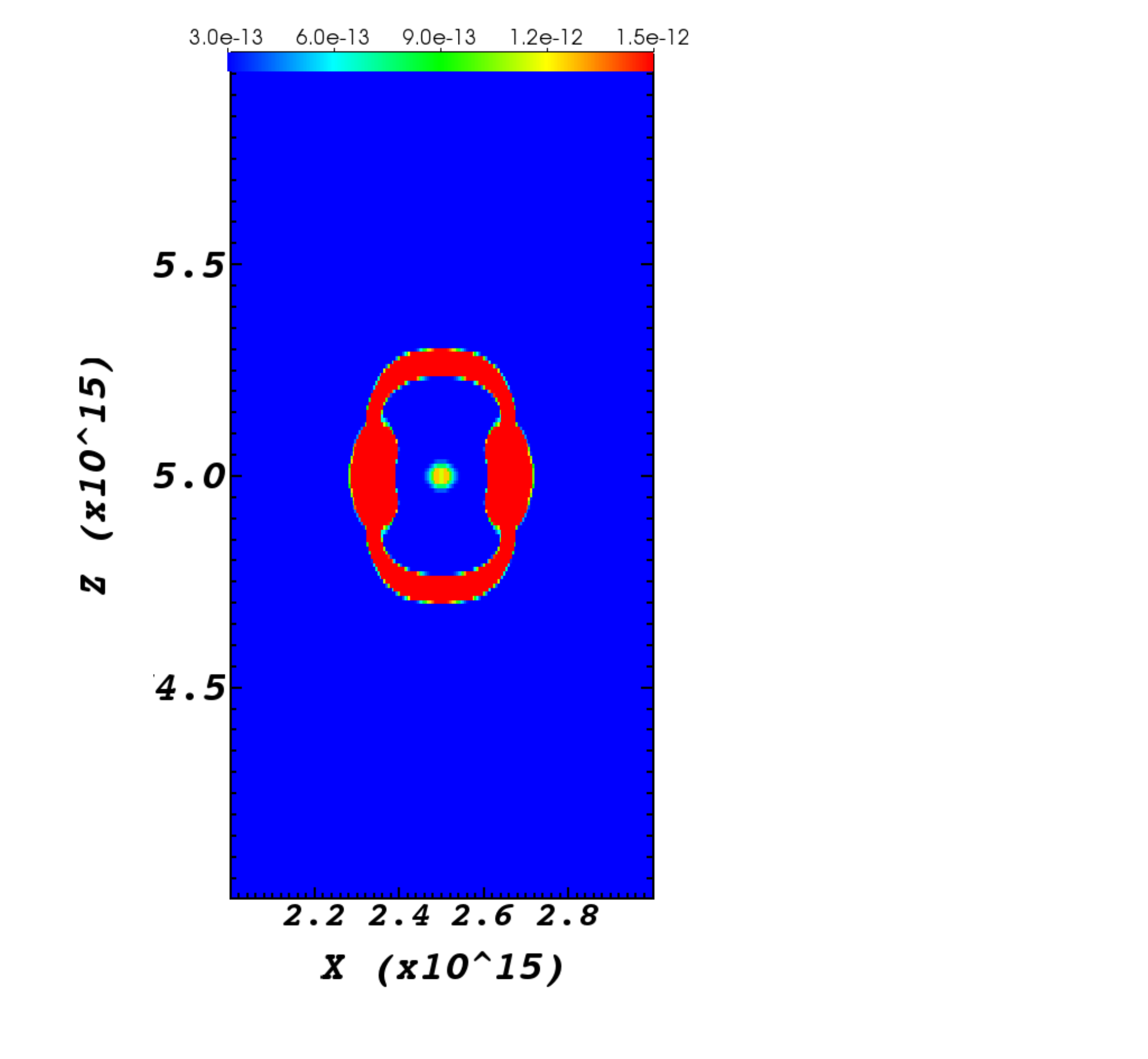}}
\subfigure[$t=0.82$~\yr]{\includegraphics[height=4in,width=4in,angle=0]{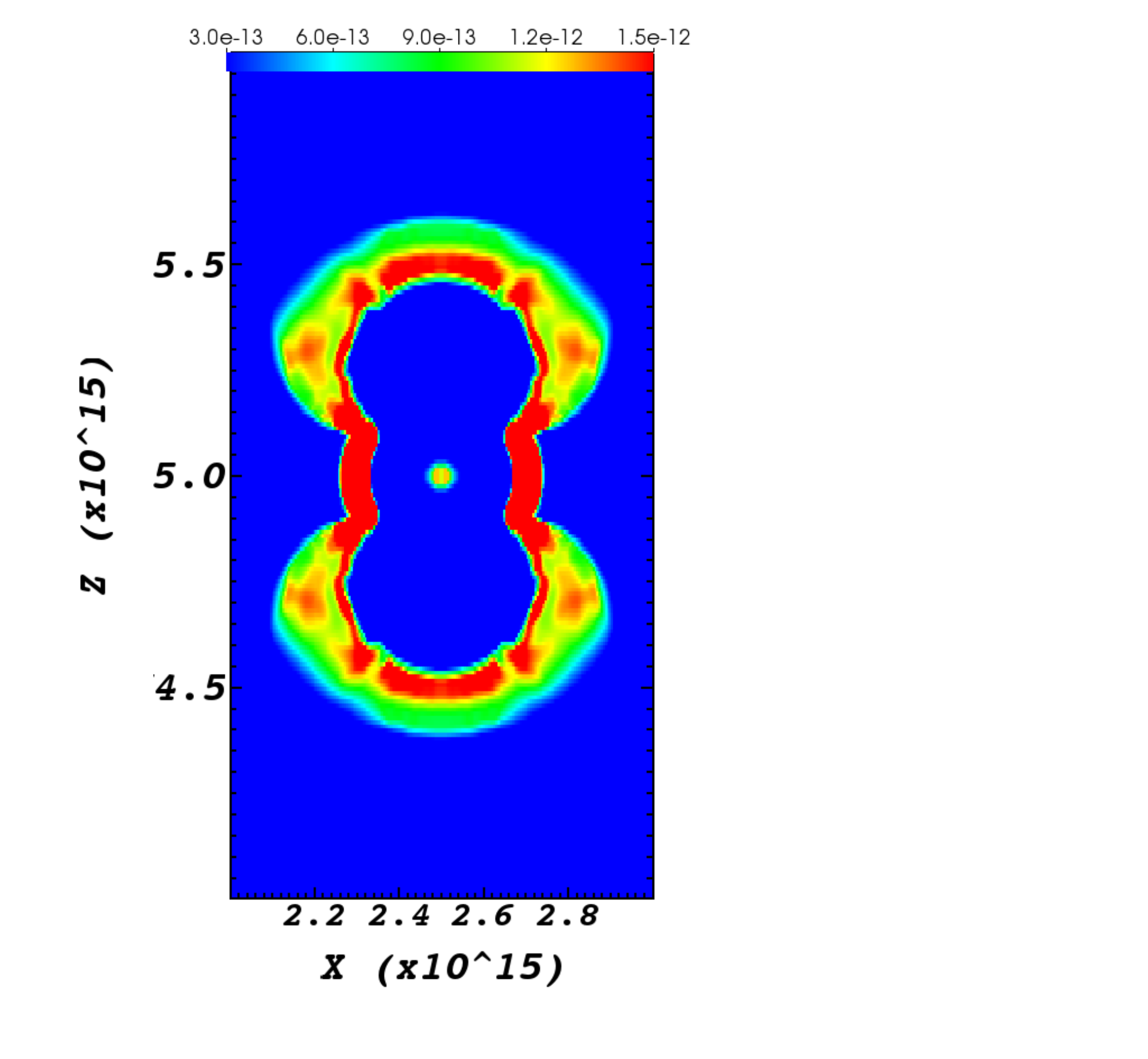}}

\subfigure[$t=1.08$~\yr]{\includegraphics[height=4in,width=4in,angle=0]{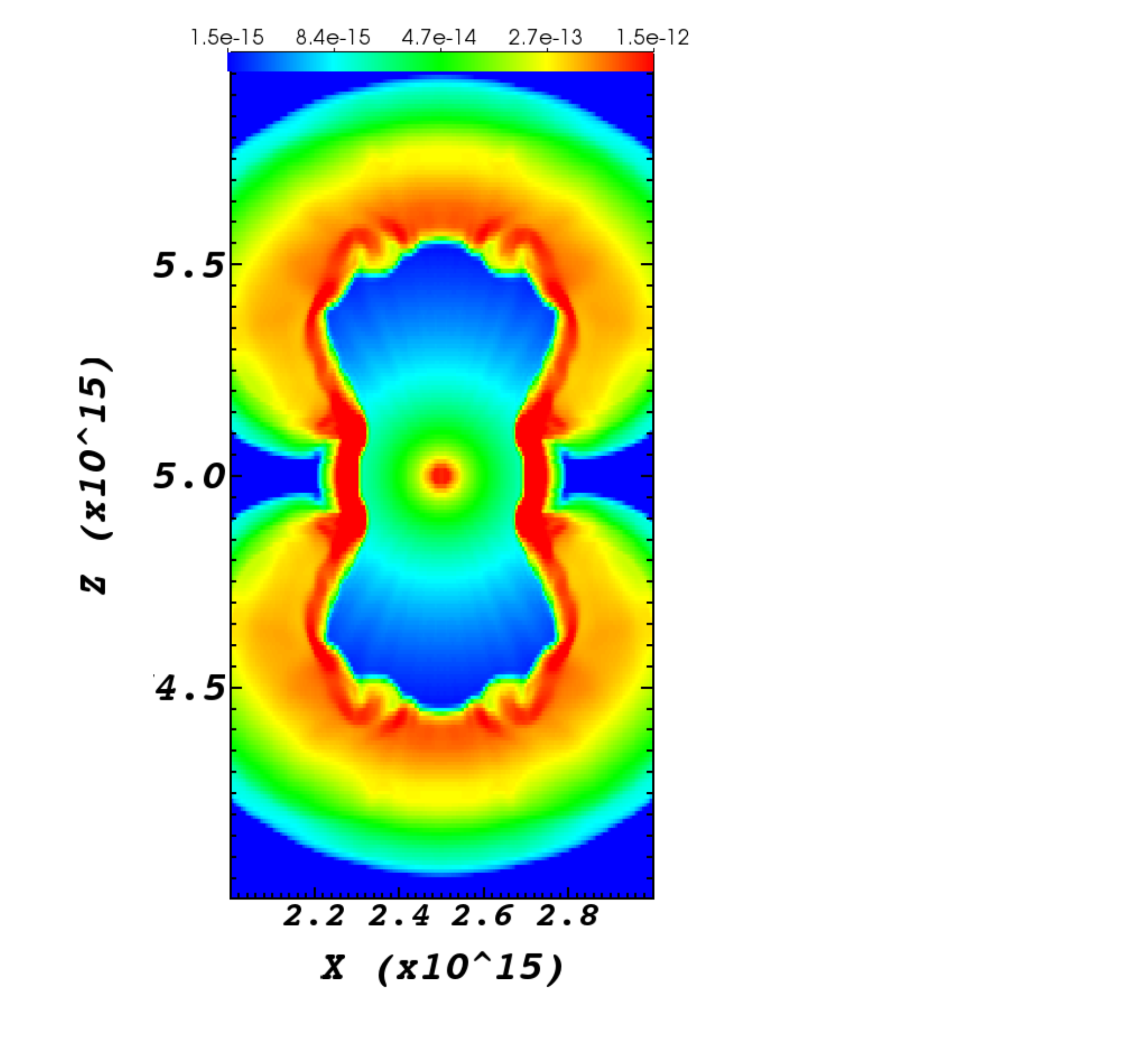}}
\subfigure[$t=1.27$~\yr]{\includegraphics[height=4in,width=4in,angle=0]{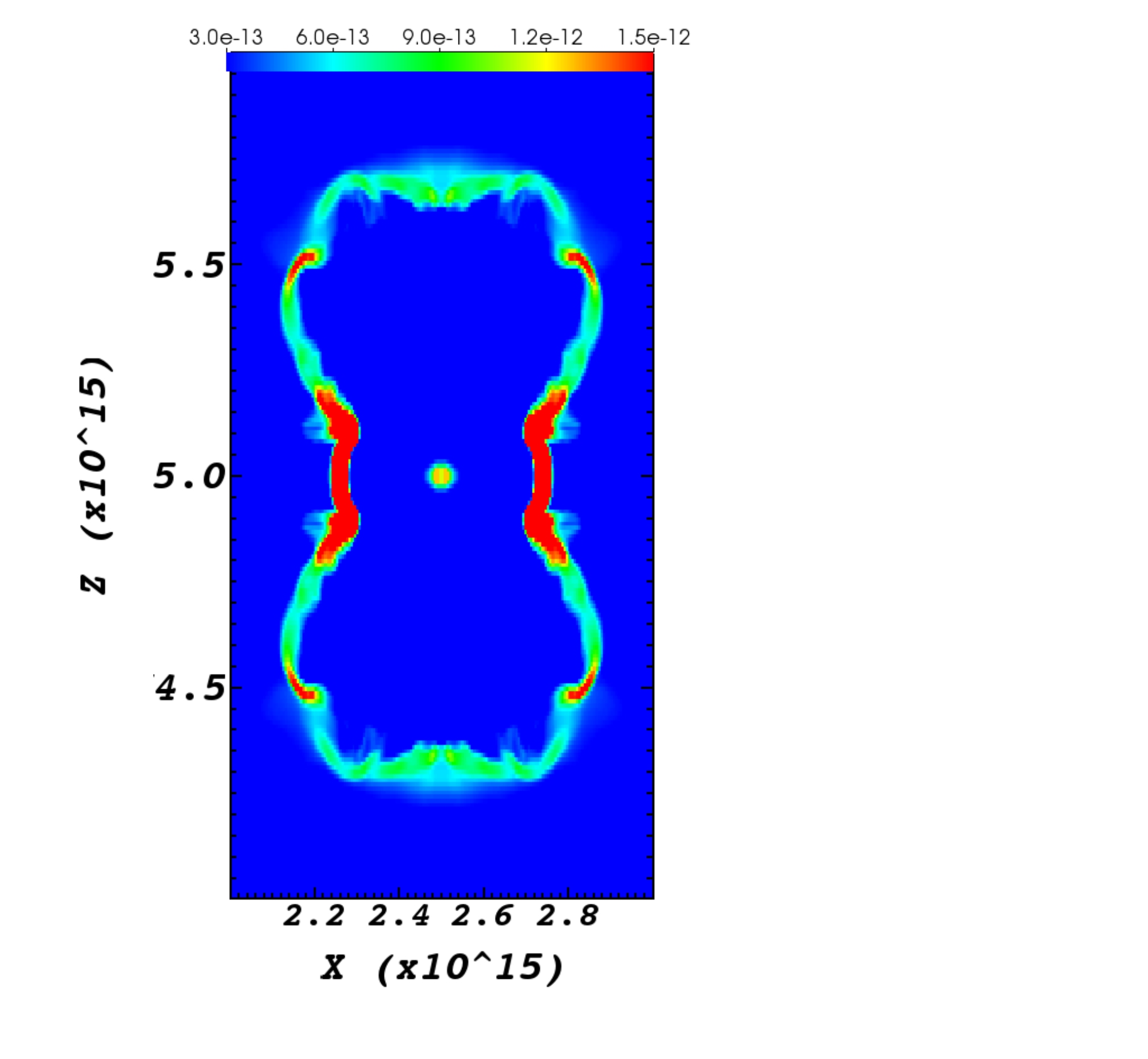}}
\caption{The density maps in the meridional plane at four times for the $\gamma=1.02$ run.
The first panel is at the time the fast spherical winds starts.
Density color coding is in units of $\g \cm^{-3}$, with a log scale in the third panel and a linear scale in the other three panels. Units on the axes are in $\cm$.
The flow that leads to the structure in the first panel is composed of two opposite jets that are injected at $r=10^{14} \cm$ along the Z axis, with a half opening angle of $50^\circ$, with an initial velocity of $700 \km \s^{-1}$, and during the time period $t=0$ to $t=58~$days.  }
 \label{fig:g1}
\end{figure}

We start with the presentation of the results for the numerical simulation with $\gamma=1.02$, that mimics radiative losses at early times \citep{AkashiSoker2013}.
In Figure \ref{fig:g1} we present the density maps in the meridional plane, i.e., the plane containing the jets' axis. 
The basic interaction of the three mass loss episodes evolves as follows. As the jets interact with the dense slow shell they inflate two opposite bubbles, with instabilities that develop in the shell (\citealt{AkashiSoker2013}). We do not present the evolution of this phase, as detailed can be seen there, but only the density at the time the fast spherical wind starts.

The fast spherical wind, which is the new addition in the present study, then interacts with the bubbles. The interaction results in a complicated flow structure and instabilities because of the shape of the bipolar structure formed by the jets. One consequence is the formation of dense rings away from the equatorial plane. The two rings can be seen as red regions (high density) away from the equatorial plane in the last panel of Figure \ref{fig:g1}: the upper ring crosses the meridional plane at that time near $(x,y)=(2.16, 5.5)$ and $(x,y)=(2.84, 5.5)$, and the lower ring crosses the meridional plane near $(x,y)=(2.16, 4.5)$ and $(x,y)=(2.84, 4.5)$, where distances are in units of $10^{15} \cm$.
There is a large dense region in the equatorial plane as well (\citealt{AkashiSoker2013}). We do not study it. As we discuss later, it is possible that the first mass loss episode, that of the dense shell, has a mass concentration at mid-latitudes, with the result of denser outer rings and a less massive equatorial ring.

\begin{figure}

\subfigure[$t=1.27$~\yr]{\includegraphics[height=2.5in,width=2.25in,angle=0]{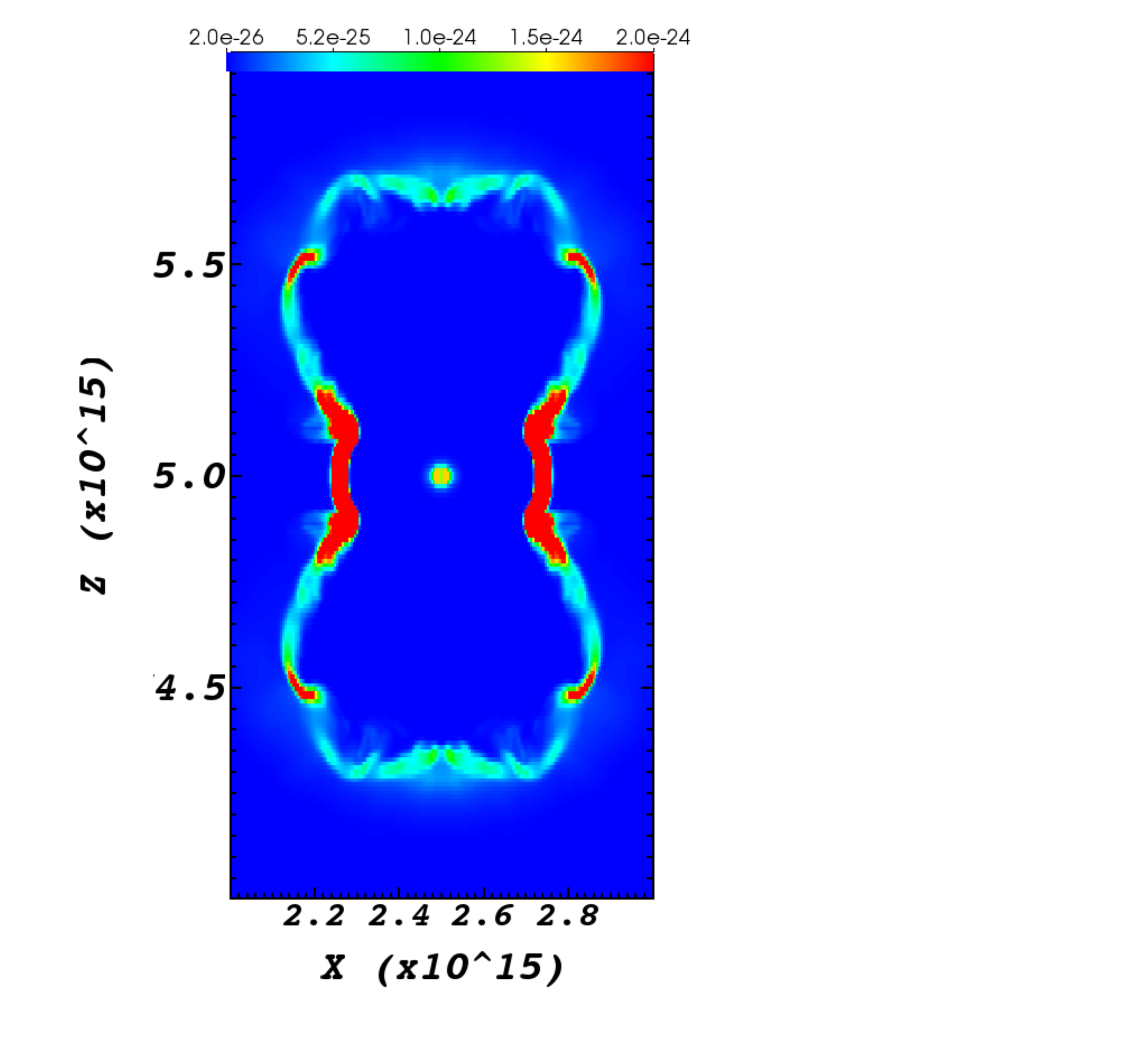}}
\subfigure[$t=1.27$~\yr  (log scale) ]{\includegraphics[height=2.5in,width=2.25in,angle=0]{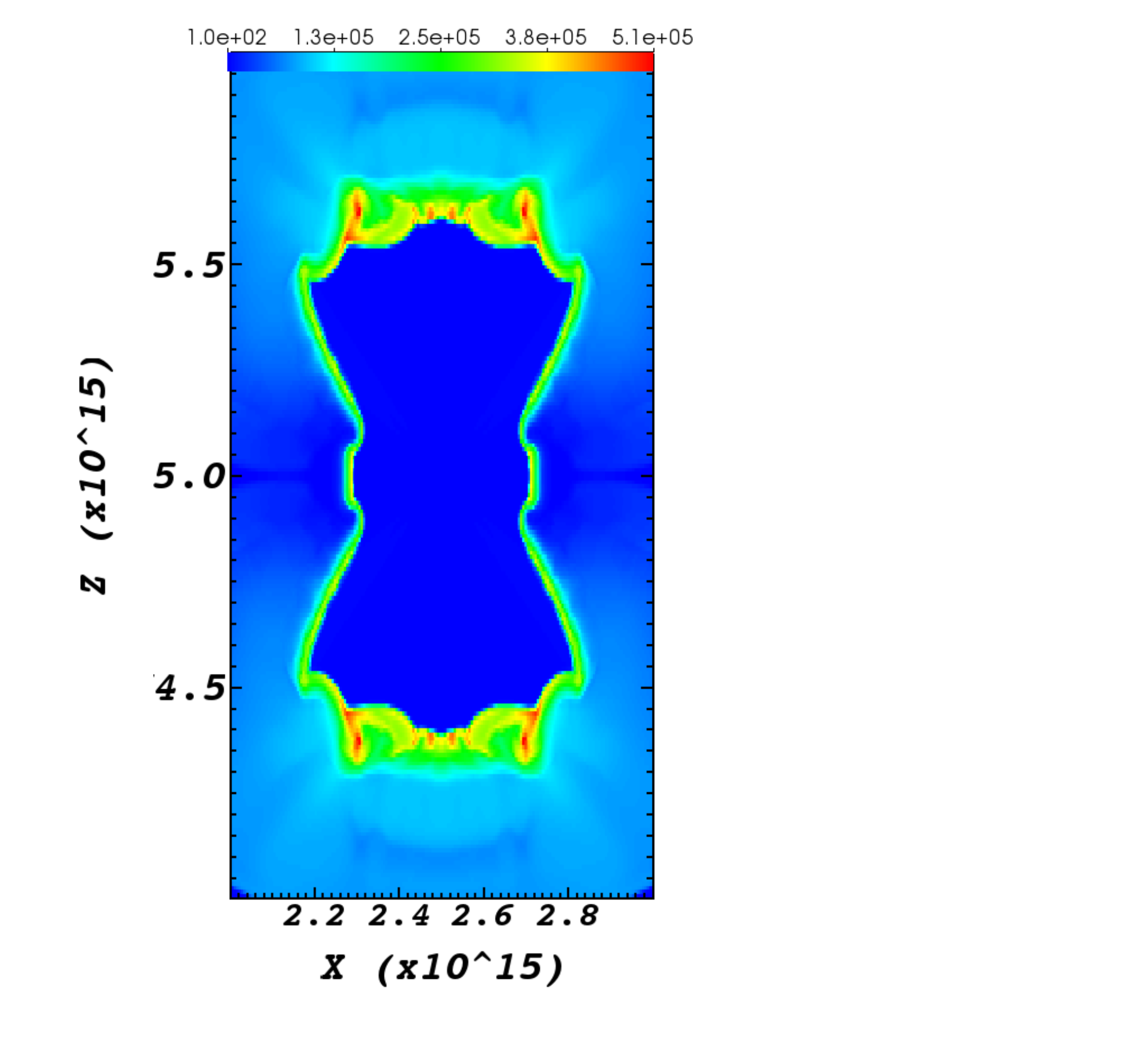}}
\subfigure[$t=1.27$~\yr]{\includegraphics[height=2.5in,width=2.25in,angle=0]{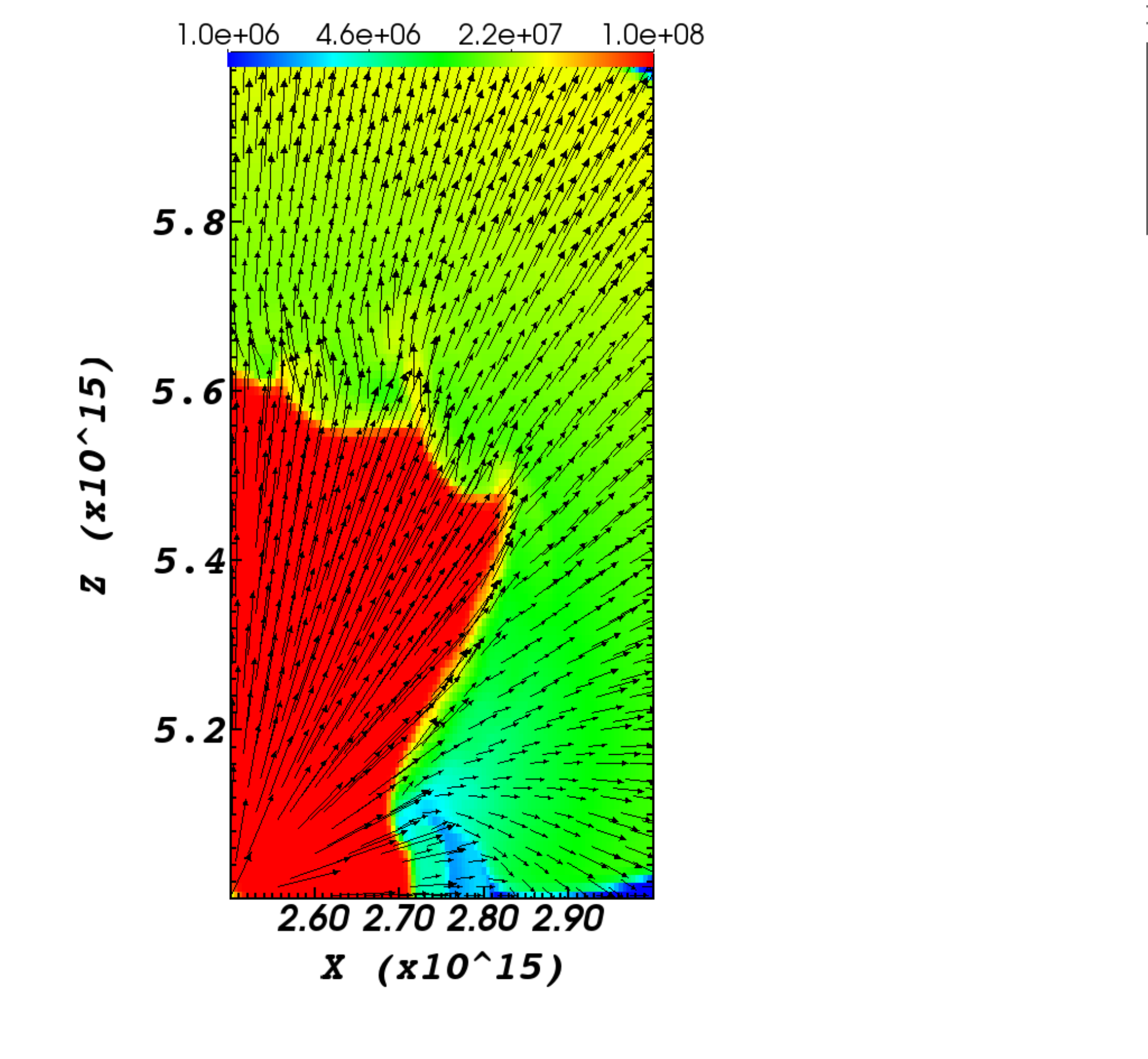}}

\caption{Some quantities in the meridional plane at time $t=1.27 \yr$ for the run presented in Figure \ref{fig:g1}.
Left: The squared of the density in $g^2 \cm^{-6}$. Middle: The temperature map in K. Right: The  velocity magnitude map (log scale in $\cm \s^{-1}$) with the velocity direction depicted by the arrows. The length of the arrows represents four velocity regimes,
$0-100 \km \s^{-1}$, $100-200\km \s^{-1}$, $200-600\km \s^{-1}$, and $600-1000\km \s^{-1}$. To better resolve the arrows we show only one quarter of the meridional plane. }
\label{fig:g2}
\end{figure}

To emphasize the double-ring structure that might be revealed in observations, we present the scaled emissivity, i.e., the density square, in the meridional plane in the left panel of Figure \ref{fig:g2}. The time of the simulation is $t=1.27 \yr$, but as mentioned above this is due to numerical limitations. In reality the time after the fast wind has shaped the nebula can last for tens to hundreds of years, and even thousands of years. In the middle panel in the same figure we present the temperature, and in the right panel the velocity of the flow. From the left panel we learn that the emissivity of the thin shell around the bubble is much higher than that of the surroundings. The rings, the two outer rings and the equatorial ring,  have a higher emissivity than the rest of the thin shell. Namely, in observations the rings will form a clear structure, but the thin shell will also be presence. We will come back to this later.

The contrast in density and emissivity between the rings and the rest of the shell we obtain is much lower than the contrast inferred in observations of the outer ring of SN~1987A. We find the density in the rings to be larger by a factor of about 2 relative to the rest of the shell (red versus green in the last panel of Figure \ref{fig:g1}. The density of the rings of SN~1987A is a factor of about 100 larger than the density of surroundings of the rings \citep{Burrowsetal1995, Panagiaetal1996}. We do not manage to account for this density ratio.

\begin{figure}

\subfigure[$t=0.82$~\yr]{\includegraphics[height=2.7in,width=2.7in,angle=0]{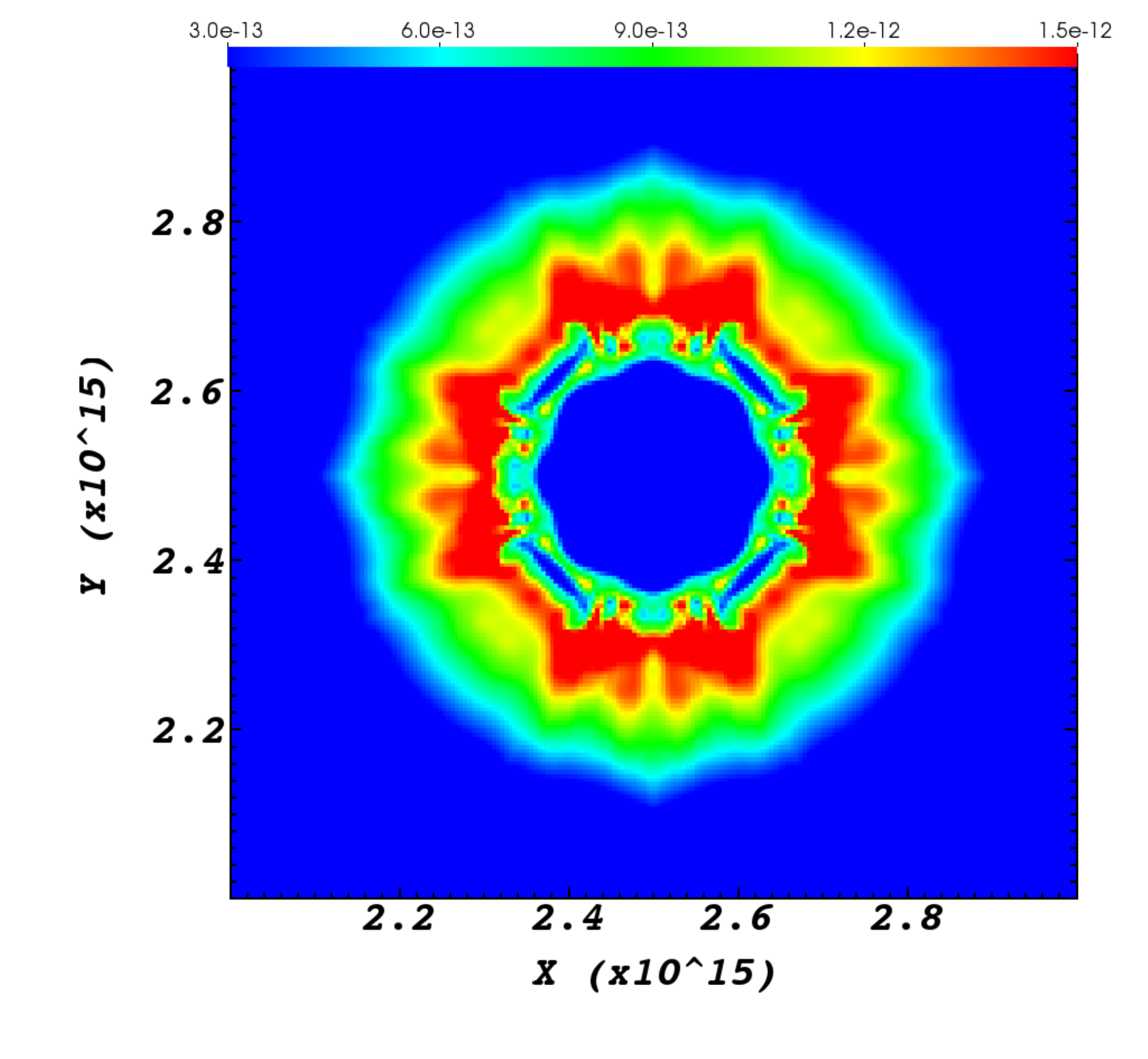}}
\subfigure[$t=1.08$~\yr]{\includegraphics[height=2.7in,width=2.7in,angle=0]{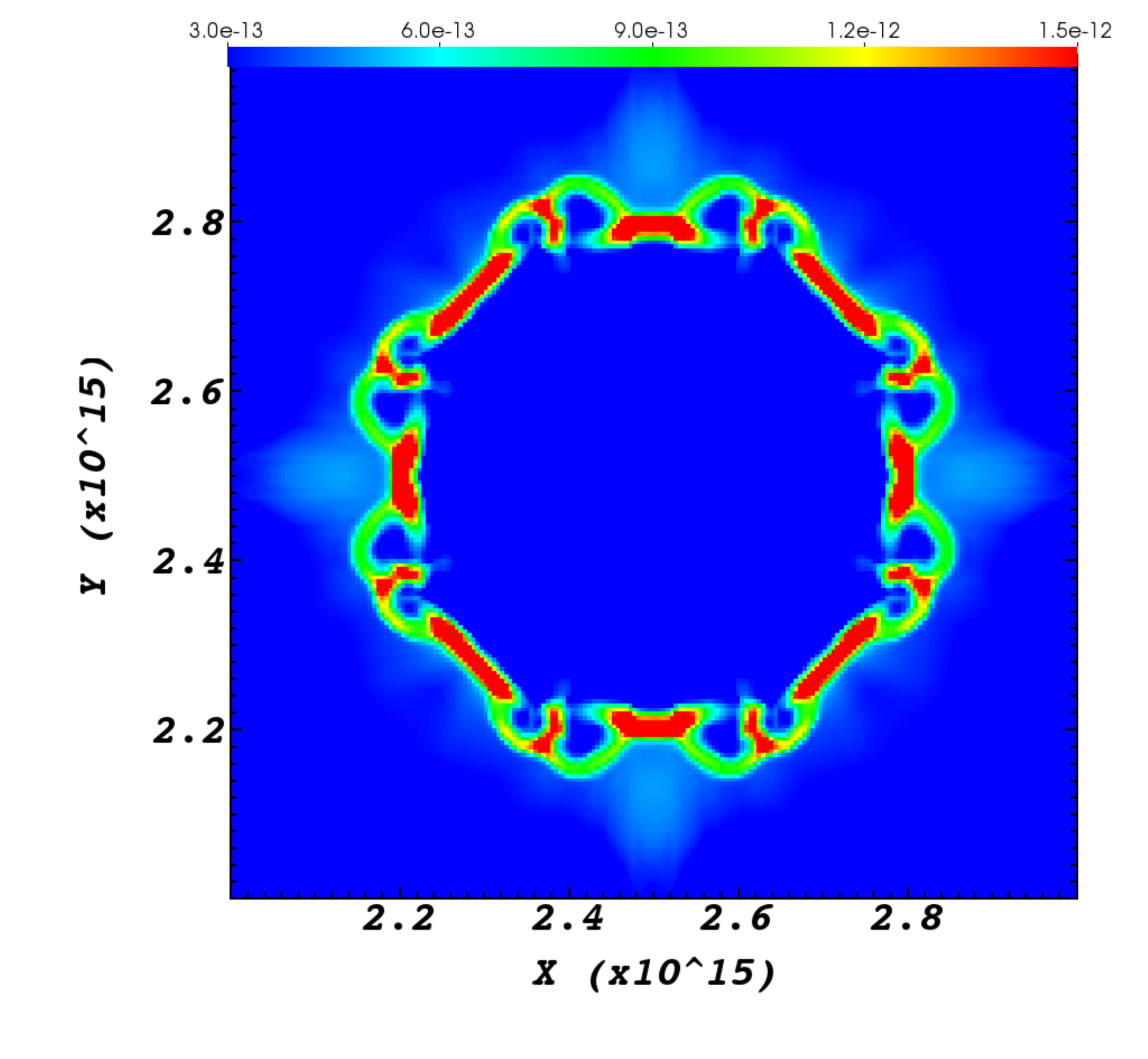}}

\subfigure[$t=1.27$~\yr]{\includegraphics[height=2.7in,width=2.7in,angle=0]{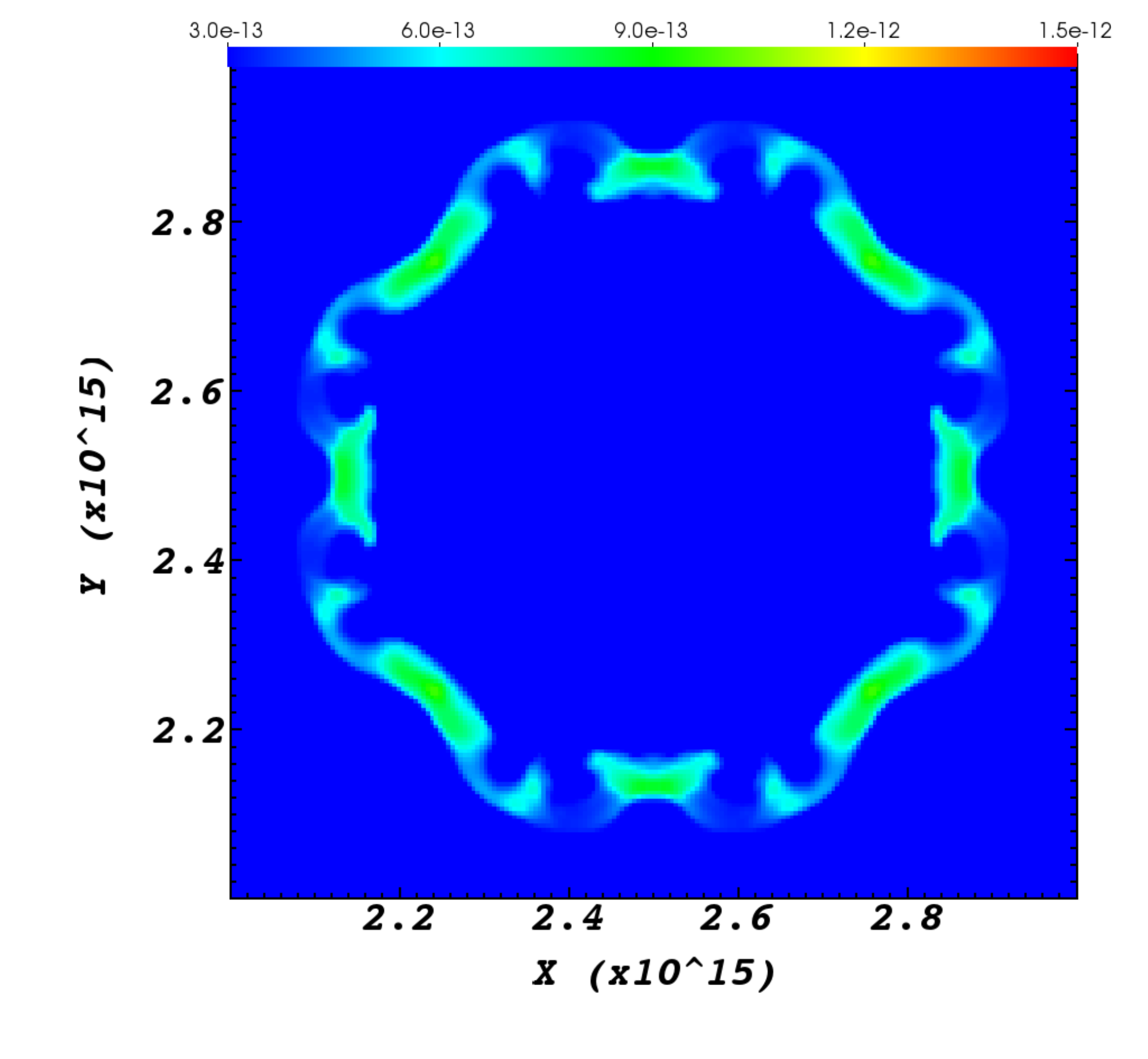}}
\subfigure[$t=1.27$~\yr]{\includegraphics[height=2.7in,width=2.7in,angle=0]{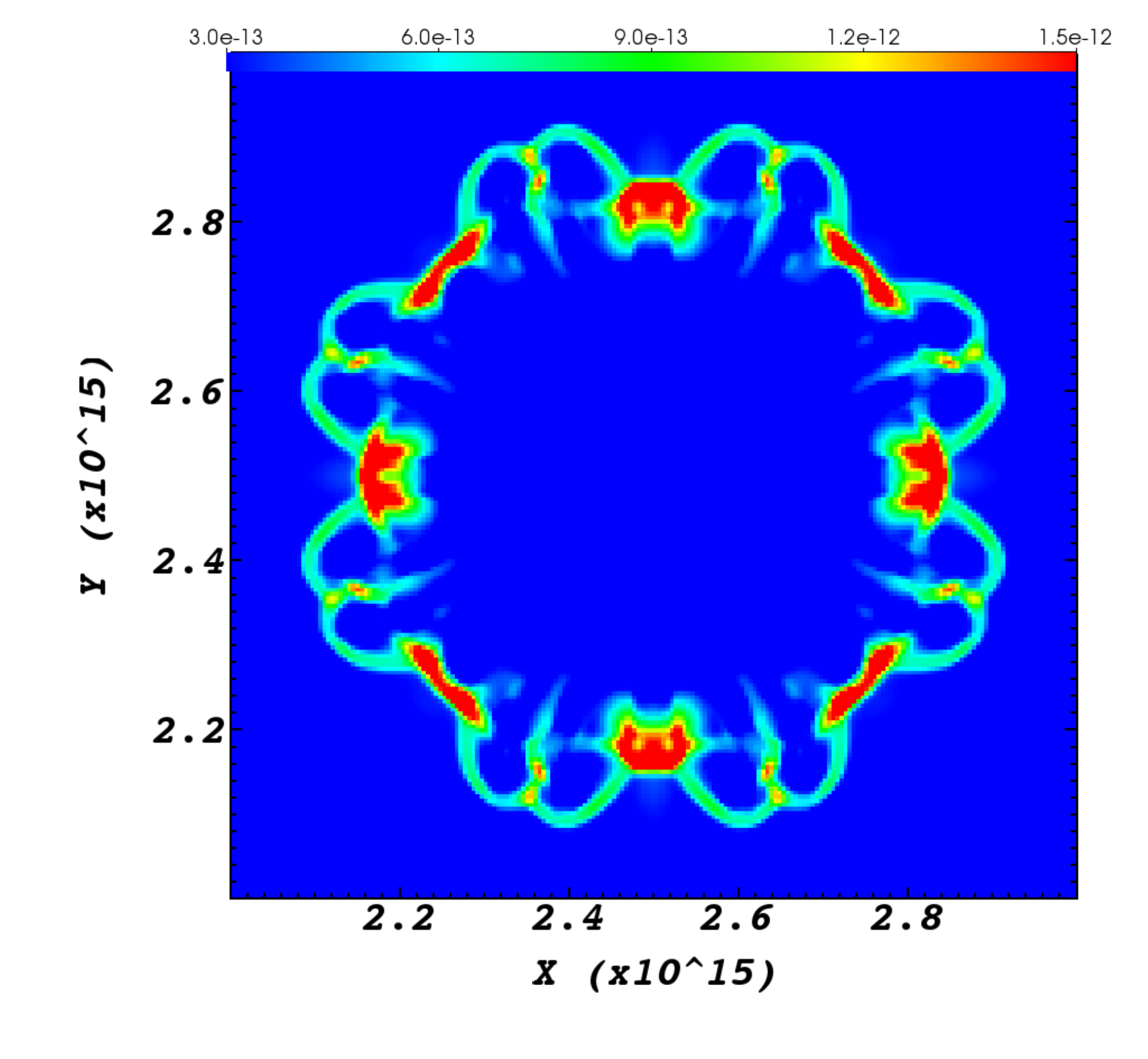}}

\caption{The density maps at 3 times of the $\gamma=1.02$ run presented in Figures \ref{fig:g1} and \ref{fig:g2}, but in planes parallel to the equatorial plane. The first three panels present maps at three times in the $z=5.4 \times 10^{15} \cm$ plane (a distance of
$0.4 \times 10^{15} \cm$ from the equatorial plane). The fourth panel is at the time  $t=1.27 \yr$ but in the $z=5.5 \times 10^{15} \cm$ plane.
Color coding is in $\g \cm^{-3}$ and units on the axes are in $\cm$. The panels emphasize the development of instabilities in the shell around the low-density bubble.
The fine detail structures of the instabilities are determined also by the numerical
grid, but their existence is physical.
 }
 \label{fig:g3}
\end{figure}

The flow becomes unstable in many place. Instabilities start from the interaction of the jets with the dense shell.  In \cite{AkashiSoker2013} we found strong instabilities on the boundaries of the bubble, and in \cite{Akashietal2015} we studies the instabilities near the equatorial plane.
Here we present the instabilities after the fast wind has interacted with the dense boundary of the bubble. In Figure \ref {fig:g3} we present the density maps in two planes parallel to the equatorial plane at three times, corresponding to the same run and last three epochs presented in Figure \ref{fig:g1}. The last panel corresponds to the plane of the upper rings. The ring might become clumpy according to our results. In reality there will be many smaller clumps, instead of several large one that are dictated by the numerical grid.

Motivated by the observations that the structure of the two outer rings of SN~1987A is displaced from the exploding star, we run simulations where the jets have been given a transverse (horizontal) velocity.
Namely, a velocity component perpendicular to the jets' axis (here it is perpendicular to the $z$ axis) has been added. Here we present results of a numerical simulation with a transverse given velocity of $v_t=100 \km \s^{-1}$ in the $x$ direction. All other parameters are as in the run presented in Figures \ref{fig:g1}-\ref{fig:g3}.
Such a velocity can be acquired from a jets' launching episode that is shorter than the orbital period. For example, consider a companion of mass $M_2=2 M_\odot$ orbiting a star of $20 M_\odot$ at $a=2 \AU$. The orbital period is $220$~day, which is about 4 times the duration of the jets' launching episode used here. The relative velocity of the two stars is about $100 \km \s^{-1}$.

\begin{figure}

\subfigure[$t=0.82$~\yr]{\includegraphics[height=2.5in,width=2.25in,angle=0]{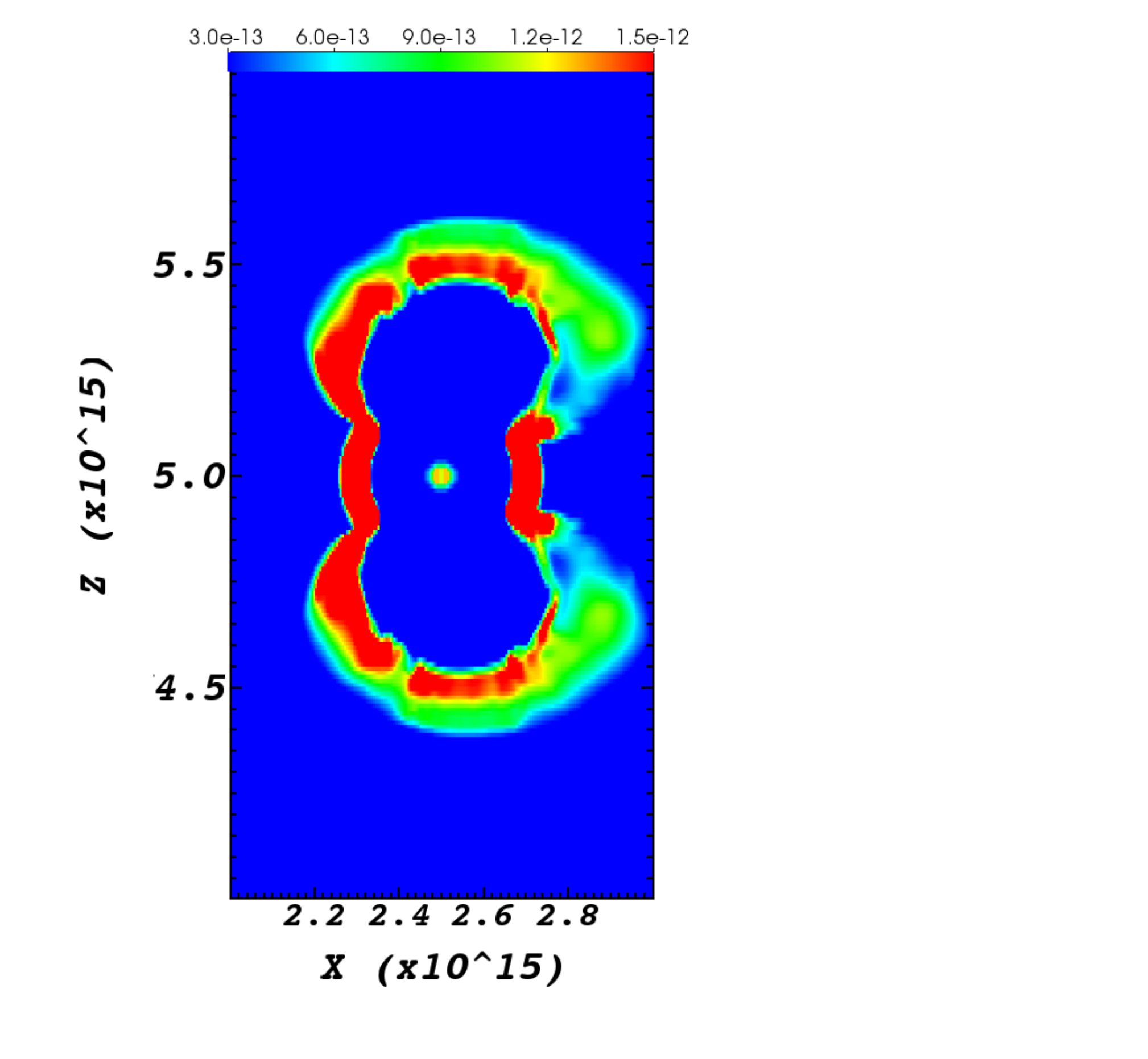}}
\subfigure[$t=1.08$~\yr]{\includegraphics[height=2.5in,width=2.25in,angle=0]{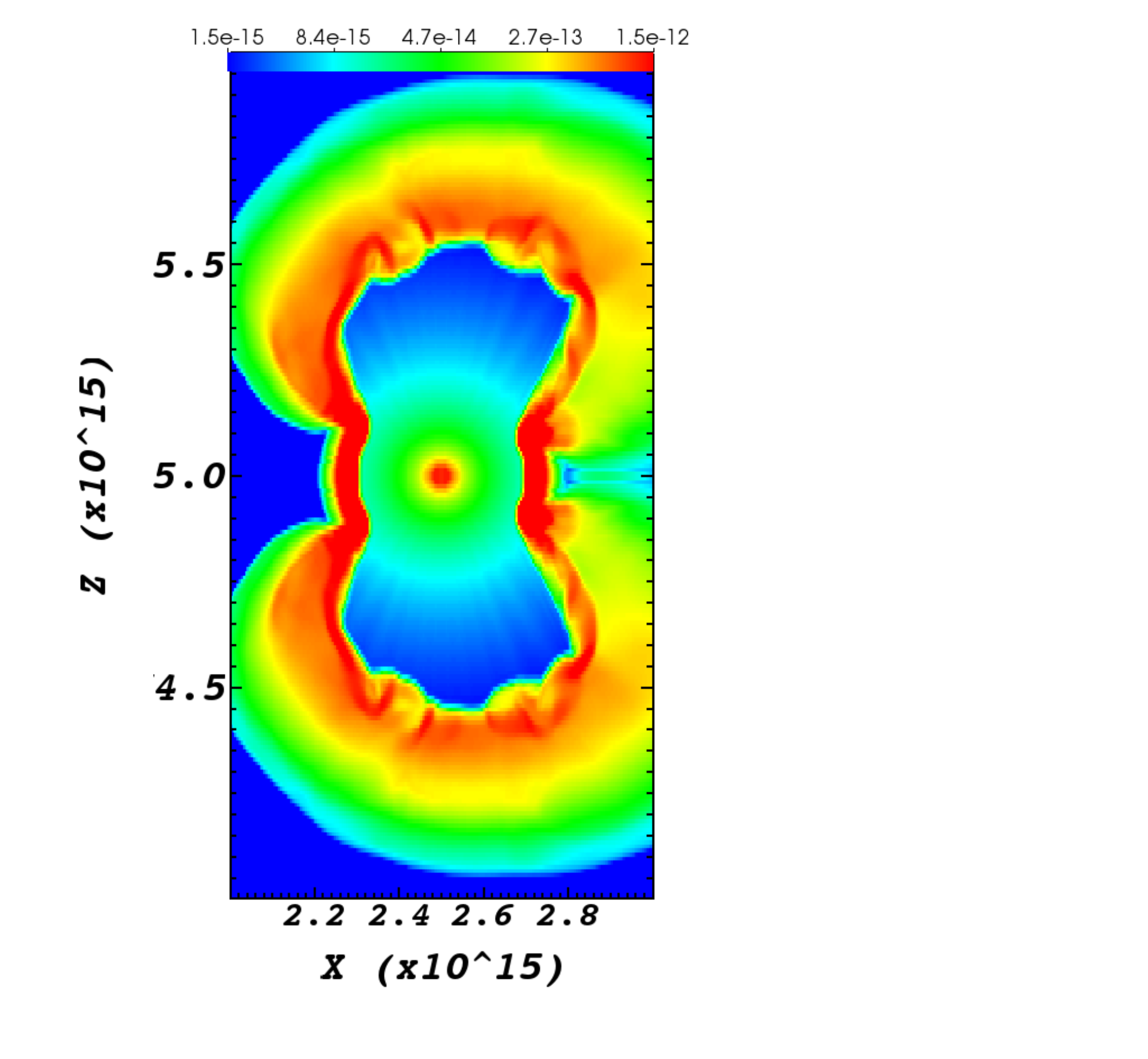}}
\subfigure[$t=1.27$~\yr]{\includegraphics[height=2.5in,width=2.25in,angle=0]{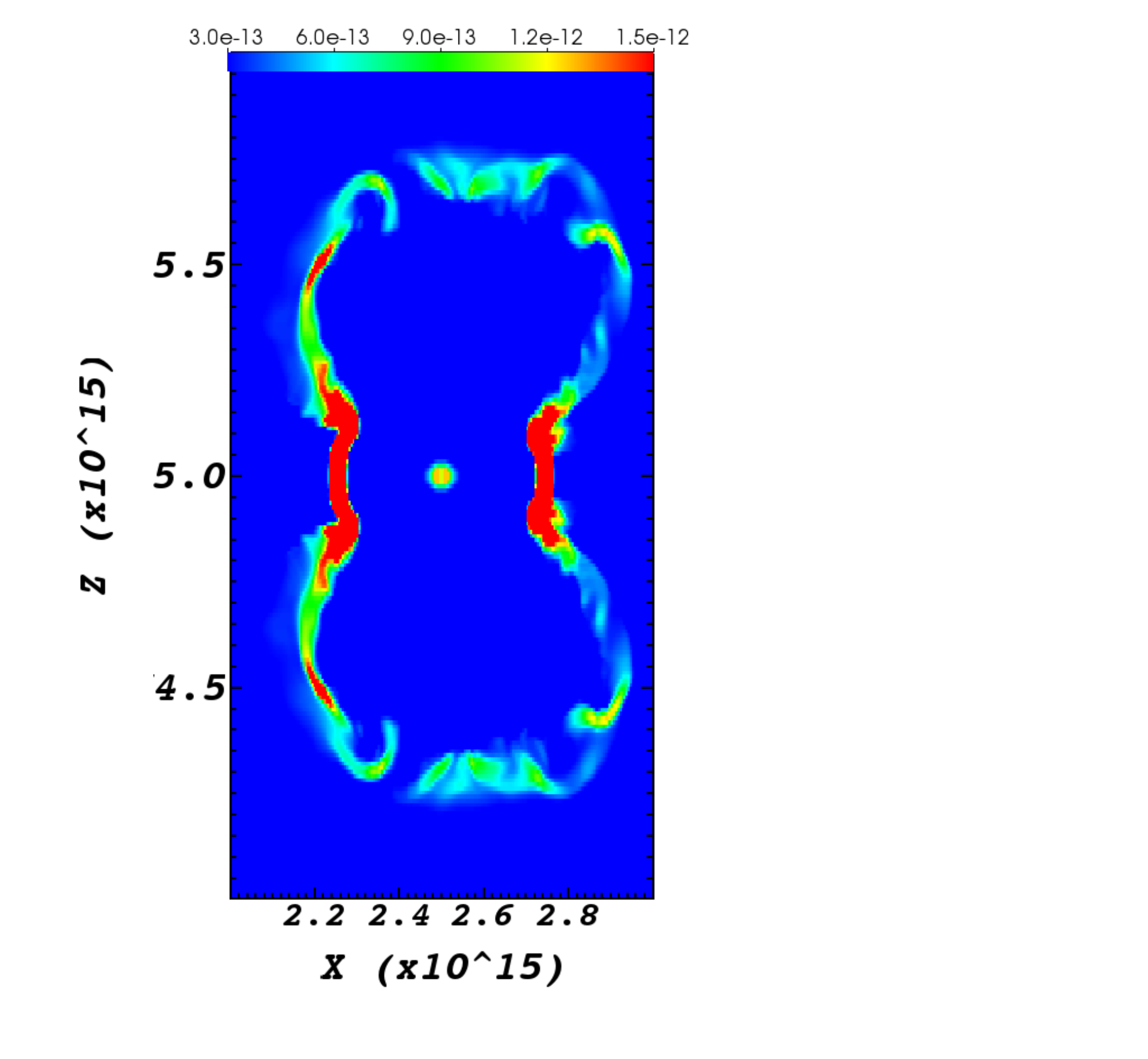}}

\caption{Same as last three panels of Fig. \ref{fig:g1}, but the jets have been injected with an initial extra velocity of  $v_t=100 \km \s^{-1}$ in the $x$ direction. Otherwise all parameters are as in the run presented there. }
 \label{fig:g4}
\end{figure}

We present the results of the displaced-jets run in Figure \ref{fig:g4}. The panels present the same times and scaling as in the last three panels of Fig. \ref{fig:g1}.
Rings are still obtained, with three characteristics of the displacement.  (1) The line connecting the center of the two outer rings is displaced from the center. (2) The more displaced side (the right side in the panels) is fainter. This fainter side is not observed in the outer rings of SN 1987A. (3) The rings are tilted, in the sense that the more displaced side of the ring was accelerated to larger distance in the $z$ direction. It seems the outer rings of SN 1987A do not show this tilt. Over all, with this additional transverse velocity given to the jets we do not reproduce the displacement observed in the outer rings of SN~1987A.

\begin{figure}

\subfigure[$t=0.82$~\yr]{\includegraphics[height=2.5in,width=2.25in,angle=0]{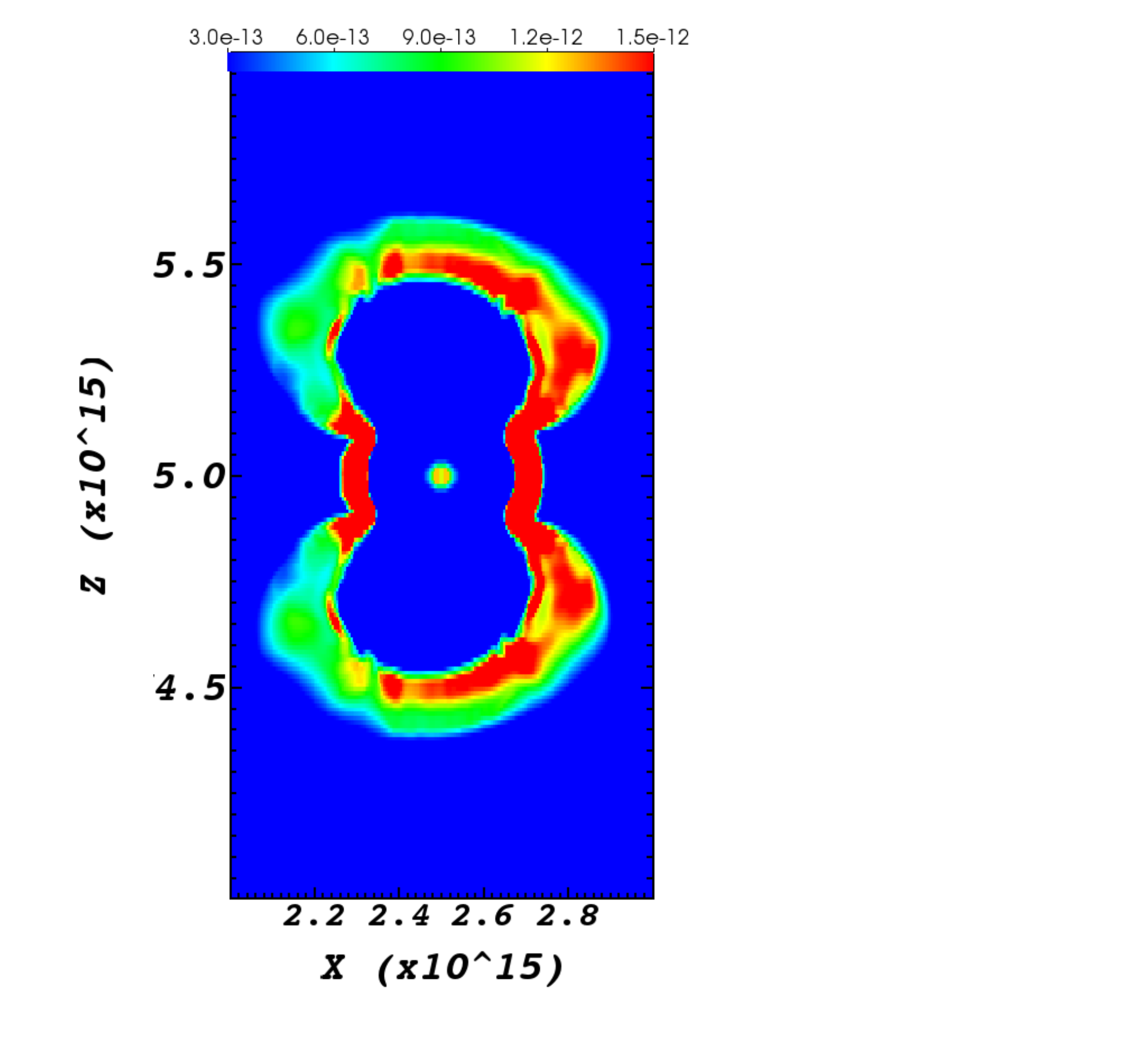}}
\subfigure[$t=1.08$~\yr]{\includegraphics[height=2.5in,width=2.25in,angle=0]{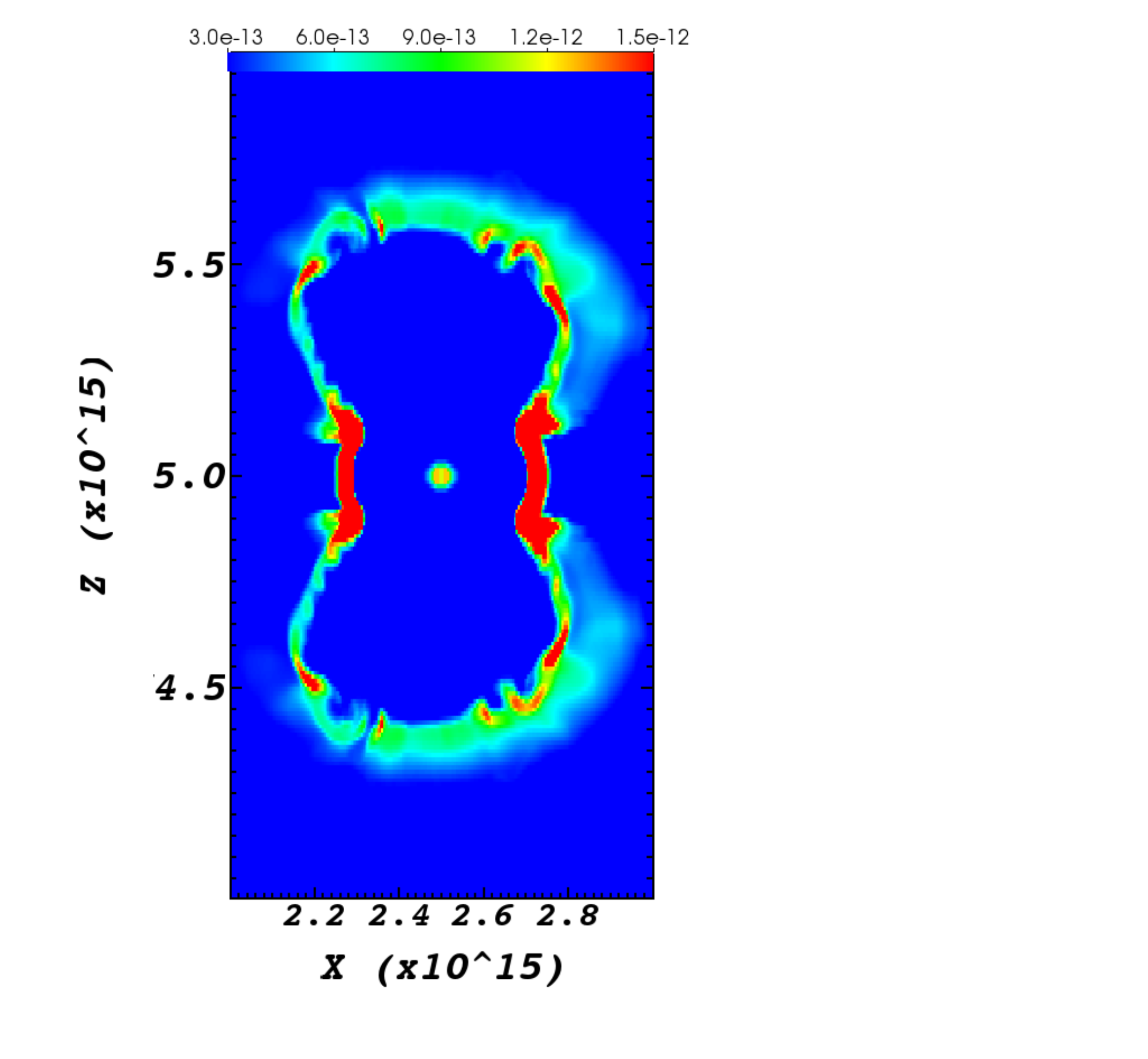}}
\subfigure[$t=1.27$~\yr]{\includegraphics[height=2.5in,width=2.25in,angle=0]{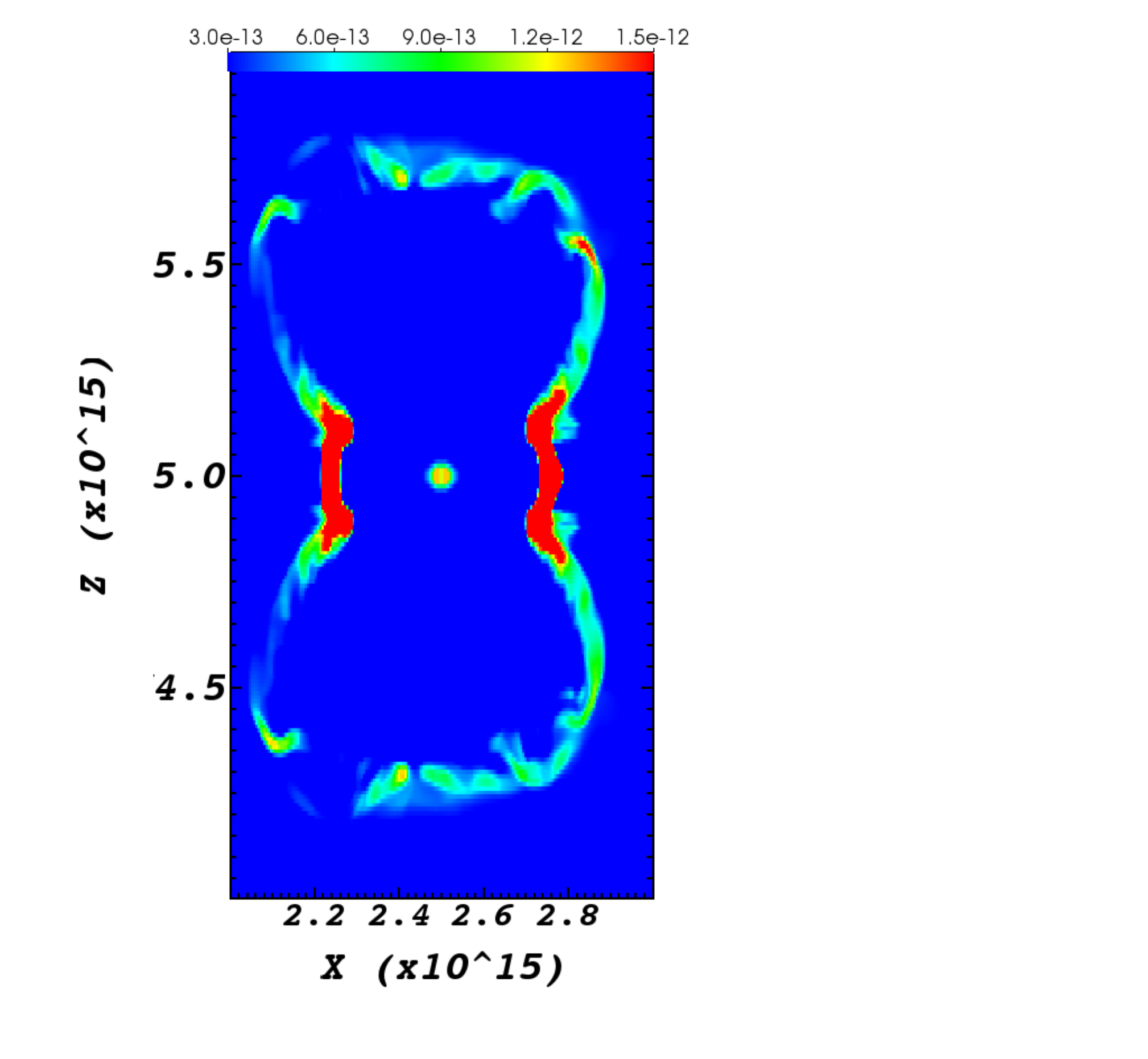}}

\caption{Same as last three panels of Fig. \ref{fig:g1}, but the spherical shell that is placed before the jets are launched is displaced in the $x$-direction by $\Delta x_s=1.5 \times 10^{13} \cm$. Otherwise all parameters are as in the run presented there. }
 \label{fig:g5}
\end{figure}

We also run one model where the spherical dense shell that was formed before the launching of the jets is displaced by $\Delta x_s=1.5 \times 10^{13} \cm$ in the $x$-direction. The jets and all other parameters as in the run presented in Figures \ref{fig:g1}-\ref{fig:g3}. The results of this displaced-shell run are presented in Figure \ref{fig:g5}. Because of this displacement there is more mass per unit solid angle in the $+x$ side (right side in the figure), and hence this side is accelerated less by the jets and the fast wind. As with the case of the displaced-jets, a double-ring system with displaced and tilted rings is formed (the outer rings). A complicate structure with clumps and one place where the fast winds almost breaks out can be seen on the front of the dense shell on each side of the equatorial plane. Here again the displaced double-ring structure is tilted and the intensity along the rings varies by a large factor. These properties are not observed in the outer rings of SN~1987A.

We can summarize this section by pointing to the double-ring system we have obtained. If jets repeat themselves, several rings might be formed. The type of flow studied here might be relevant to some bipolar PNe, but we could not reproduce the properties of the outer rings of SN~1987A.

\section{SLOW COOLING: FORMING MULTIPLE-LOBES}
\label{sec:lobes}

\begin{figure}
\subfigure[$t=0.45$~\yr]{\includegraphics[height=4in,width=4in,angle=0]{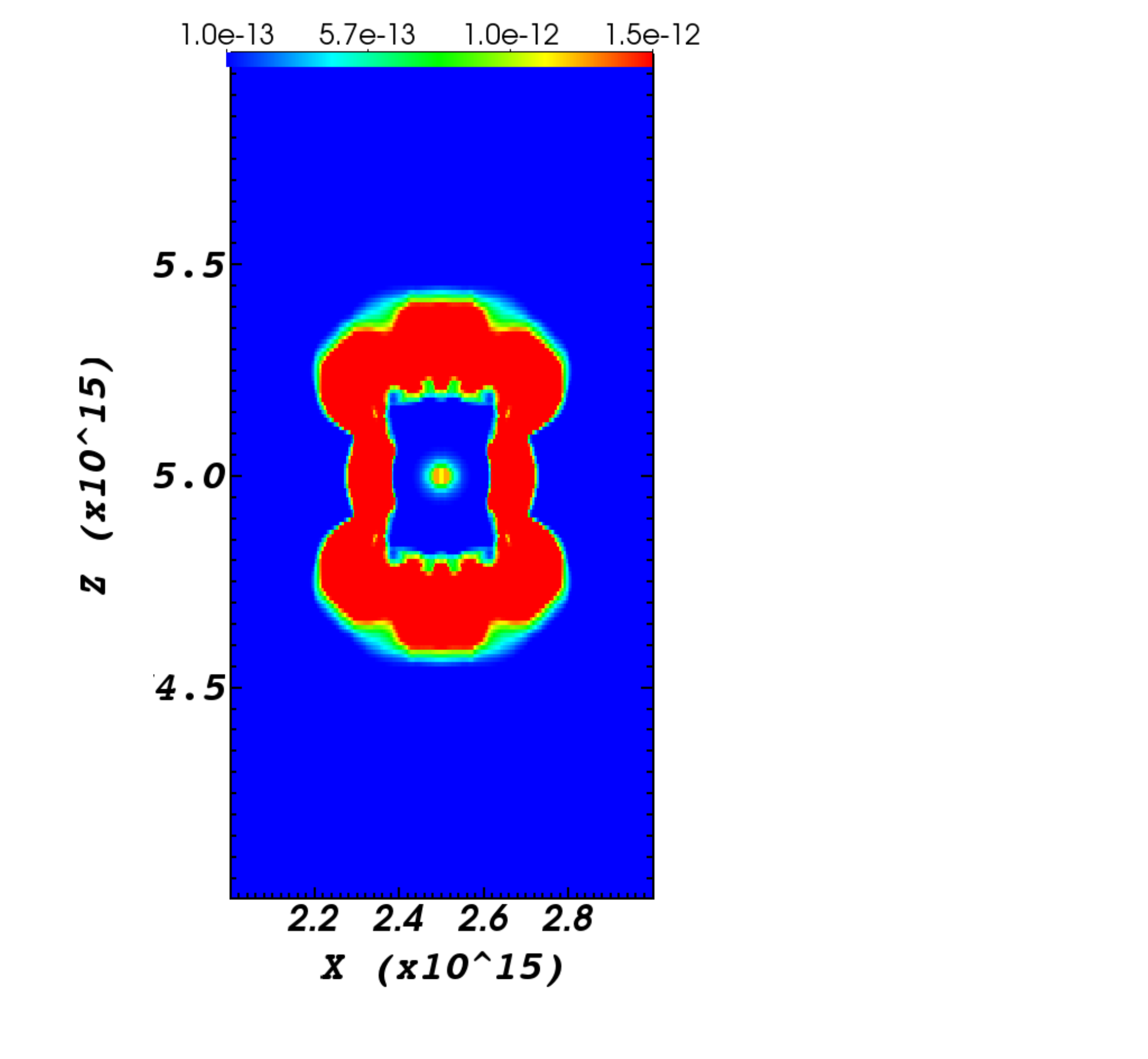}}
\subfigure[$t=0.82$~\yr]{\includegraphics[height=4in,width=4in,angle=0]{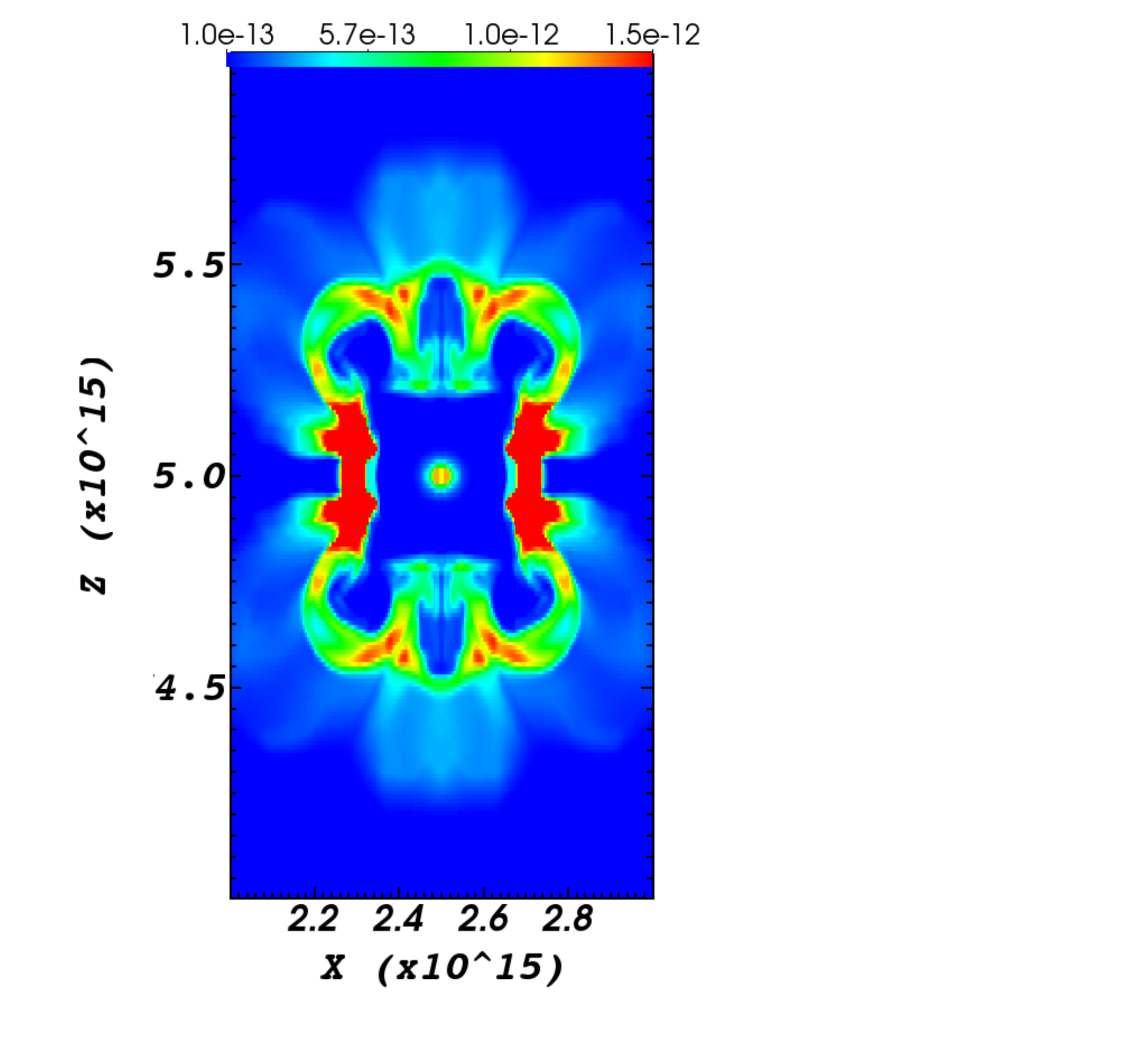}}
\subfigure[$t=1.08$~\yr]{\includegraphics[height=4in,width=4in,angle=0]{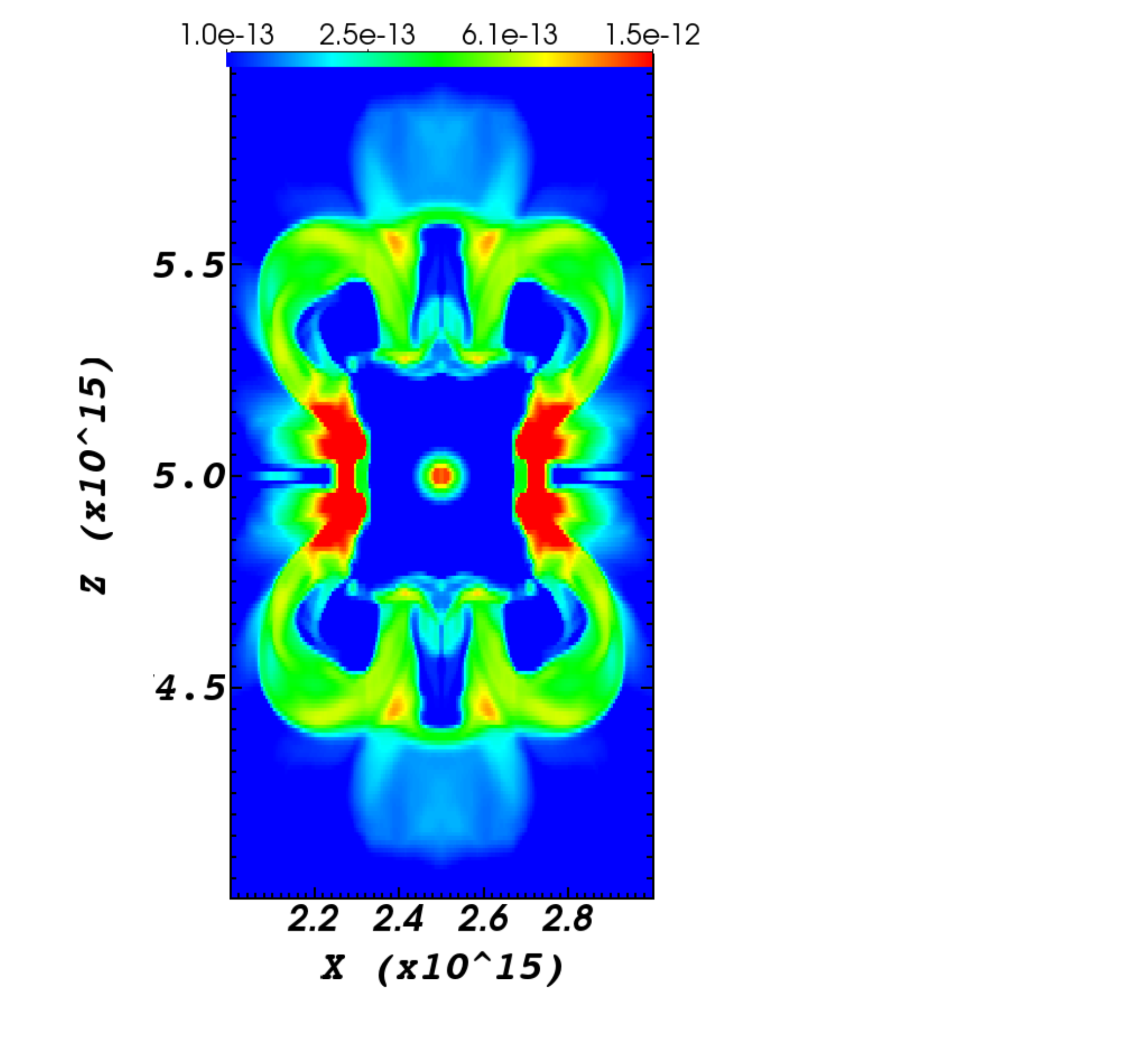}}
\subfigure[$t=1.27$~\yr]{\includegraphics[height=4in,width=4in,angle=0]{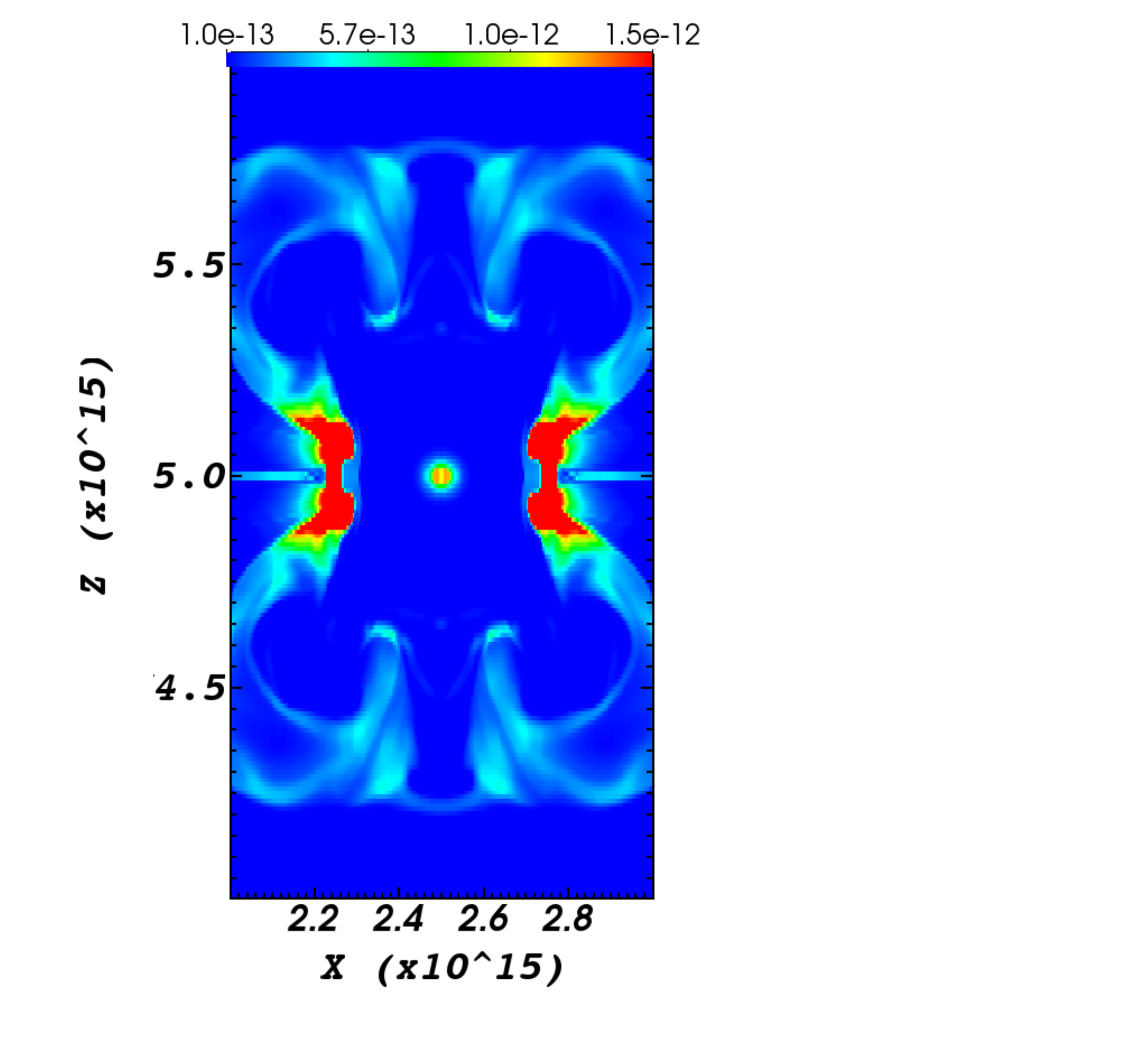}}

\caption{The density maps in the meridional plane as in Figure \ref{fig:g1}, but taking the adiabatic index to be $\gamma=1.1$, and at different times as indicatred. Other parameters of the simulation are as in the run presented in Figures \ref{fig:g1}-\ref{fig:g3}.
Density color coding is in units of $\g \cm^{-3}$ and in a linear scale. Units on the
axes are in $\cm$. }
 \label{fig:g6}
\end{figure}

\begin{figure}
\subfigure[$t=0.61$~\yr]{\includegraphics[height=4in,width=4in,angle=0]{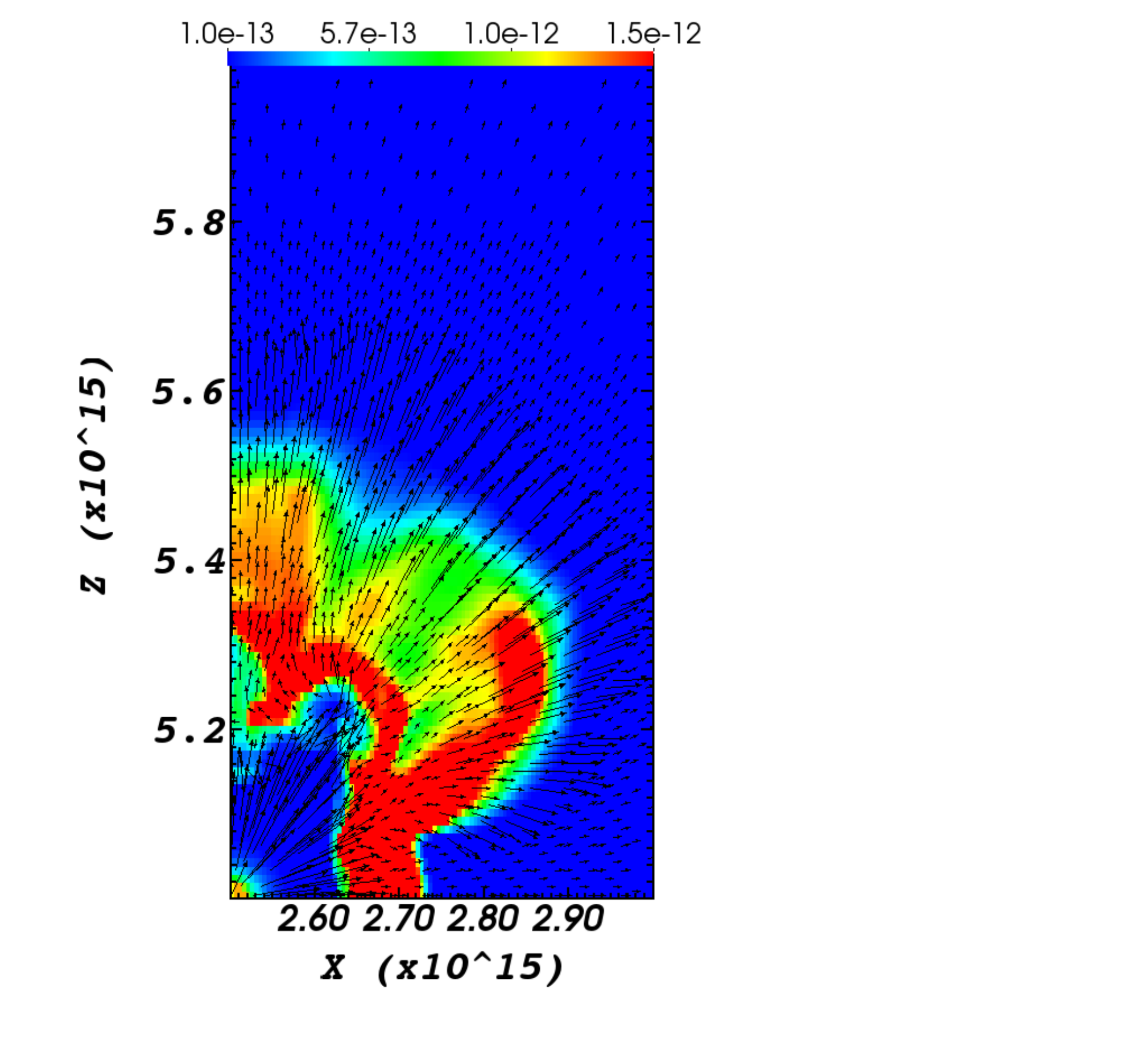}}
\subfigure[$t=0.82$~\yr]{\includegraphics[height=4in,width=4in,angle=0]{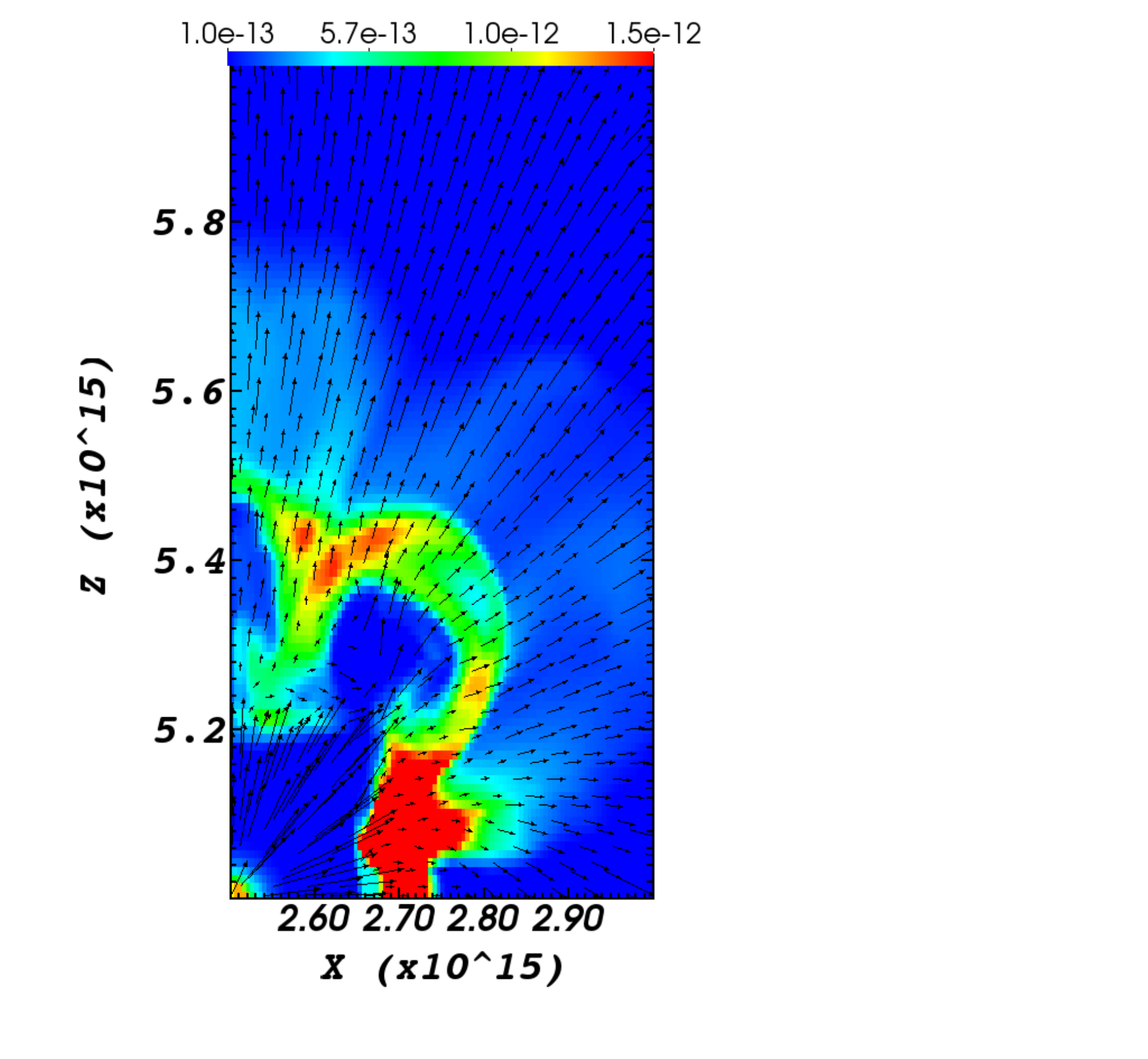}}
\subfigure[$t=1.08$~\yr]{\includegraphics[height=4in,width=4in,angle=0]{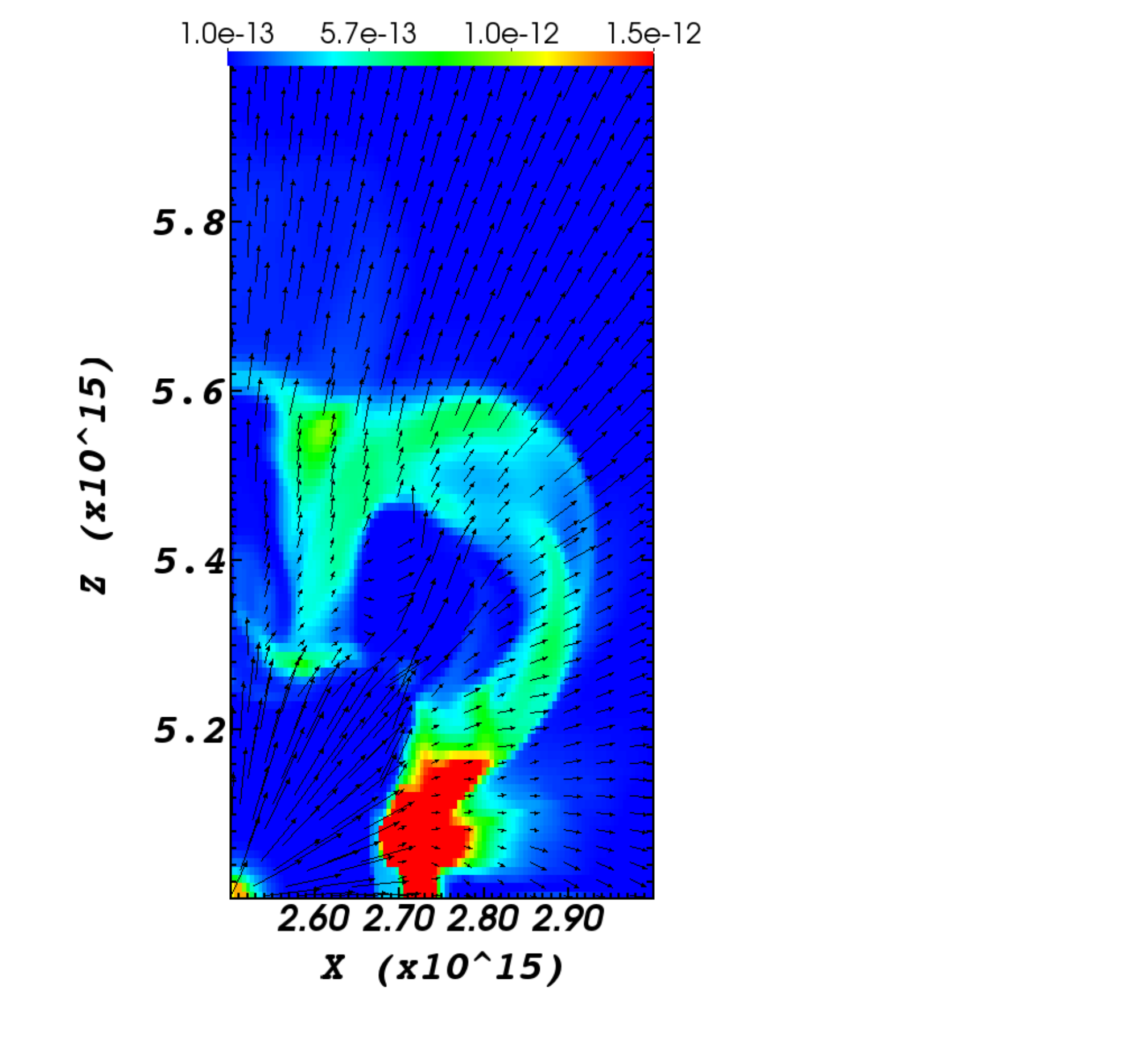}}
\subfigure[$t=1.27$~\yr]{\includegraphics[height=4in,width=4in,angle=0]{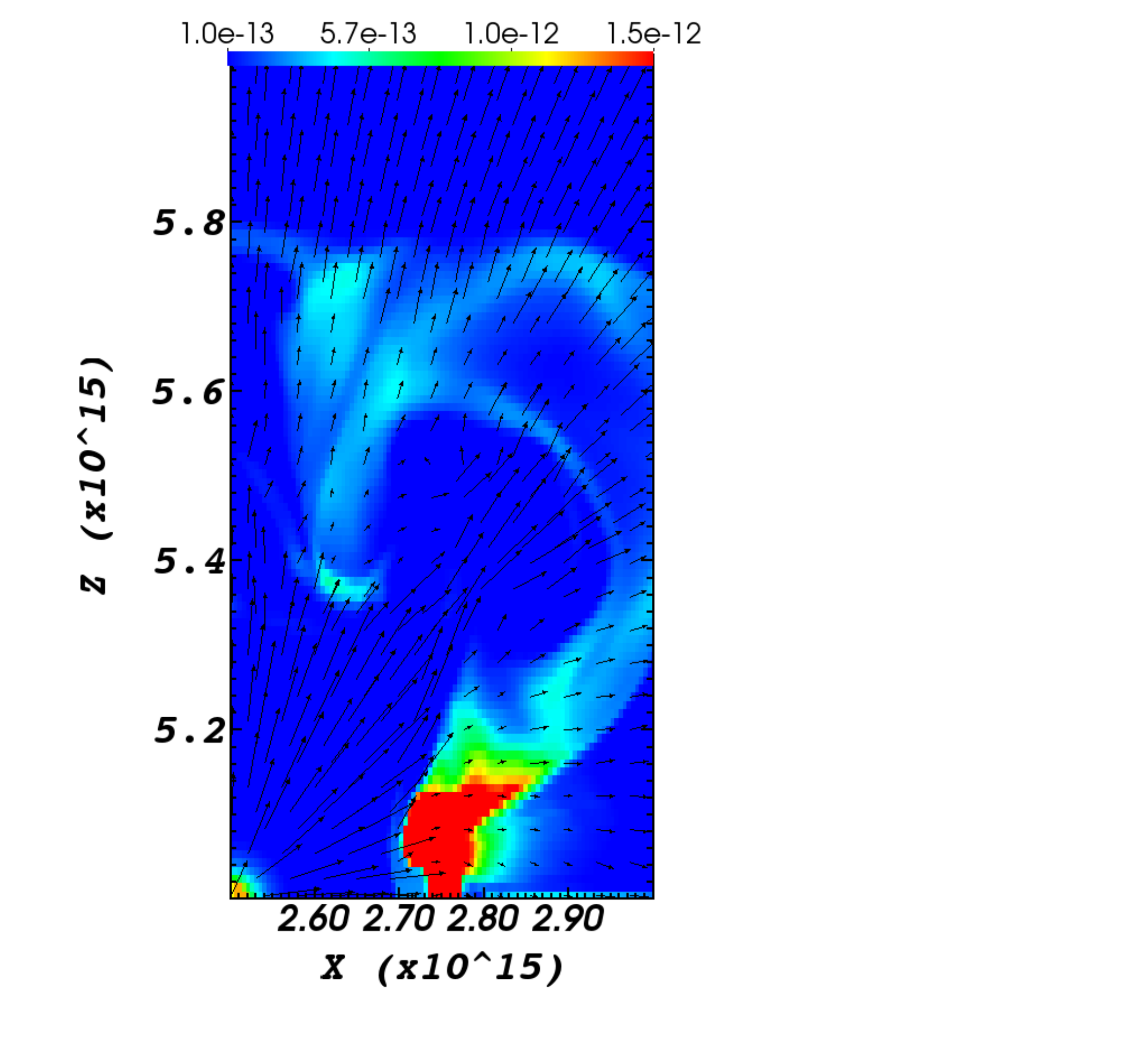}}

\caption{The density maps and velocity arrows at four times corresponding to the four panels presented in Figure \ref{fig:g6}. The length of the arrows represents four velocity regimes,
$0-100 \km \s^{-1}$, $100-200\km \s^{-1}$, $200-600\km \s^{-1}$, and $600-1000\km \s^{-1}$.
Only one quarter of the meridional plane is shown.
 }
 \label{fig:g7}
\end{figure}

We present results of a run where we set the adiabatic index to be $\gamma=1.1$ instead of $\gamma=1.02$ as in the runs presented in section \ref{sec:rings}. This represents a case where the radiative losses are moderate. All other parameters are as in the run presented in Figures \ref{fig:g1}-\ref{fig:g3}.
In this case a larger fraction of the kinetic energy of the jets and the fast wind is transferred to thermal energy of the shocked gas, and a large hot bubble is formed, as can be seen in the density and temperature maps presented in Figures \ref{fig:g6} and \ref{fig:g7}.

We do not obtain rings outside the equatorial plane, but rather a structure of 3 bubbles on each side of the equatorial plane is formed in the meridional plane.
In 3D, the structure is of a central bubble and a torus-like bubble around it. In observations, the structure will be of three pairs of bipolar lobes.
We are not aware of any PN that can be explained by this specific structure. This structure might be transients.

There are two main conclusions that we draw from this run and the one presented in section \ref{sec:rings}.
 (1) The nebular structure that is obtained from the three mass loss episodes is very sensitive to the radiative losses during the interaction.
 (2) Jets can enrich the variety of morphological structures that result from binary interaction.

\section{DISCUSSION AND SUMMARY }
\label{sec:summary}

This paper is another one in our series of papers aiming at exploring the role of jets in shaping some nebulae around evolved stars \citep{AkashiSoker2008a, AkashiSoker2008b, Akashietal2015}. The new ingredient we have added is a fast spherical wind blown after the shaping by jets. Such a wind is expected to last tens to thousands of years. Due to numerical limitations, we set the fast wind to be very short and to have a very high mass loss rate.

We mimicked radiative cooling by using a low ($\gamma=1.02$; Figs. \ref{fig:g1}-\ref{fig:g5}) and moderate ($\gamma=1.1$; Figs. \ref{fig:g6}-\ref{fig:g7}) adiabatic indices.
When cooling is rapid enough, we obtained a structure of three rings, that are seen as in red color in the bottom-right panel of Fig.\ref{fig:g1}: a massive ring in the equatorial plane, and two outer rings, one at each side of the equatorial plane.
Each outer ring ia part of a dense shell on the boundary of the bubble. The rings are much denser than regions outside and inside this shell, but not much denser than gas in the shell.
As such, our model with the parameters used cannot reproduce the outer rings of SN~1987A that are much denser than their surroundings.

The line connecting the two outer rings of SN~1987A is displaced with respect to the star that exploded.
We have tried to reproduce the displacement of the rings by displacing the spherical shell and by adding transverse velocity to the jets. These did not reproduce all properties of the outer rings.

\cite{MorrisPodsiadlowski2009} could produce many of the properties of the outer rings without jets. They rather assumed that more mass was lost at a mid latitude at each side of the equatorial plane. This is based on their study of the merger process of two stars \cite{MorrisPodsiadlowski2006}.
It is possible that the formation of outer rings with a large density contrast requires that there will be a high mass loss rate at mit-latitude, as assumed by \cite{MorrisPodsiadlowski2007} and \cite{MorrisPodsiadlowski2009} for the progenitor of SN~1987A, and as \cite{Smith2006} found for the Great Eruption of $\eta$ Carinae that formed its bipolar nebula, the Homunculus.
Our model includes three mass loss episodes: a spherical shell, jets, and a spherical fast wind.
The next step should be to replace the spherical shell by a shell with enhanced mass loss at mid-latitudes.

We could produce outer rings that might be applicable in some PNe when radiative cooling in our model was significant, as we modeled by taking $\gamma=1.02$. When the radiative losses are moderate, as we simulated by taking $\gamma=1.1$, outer rings are not formed. Instead a complicated structure of lobes arises. Further comparison with observed specific PNe requires to tune the parameters, which is beyond our scope.

Our main result is in further emphasizing the large variety of bipolar morphological structures around evolved stars that can be shaped by jets, and in particular to attributing a role to the fast wind.

\bigskip
{\bf ACKNOWLEDGEMENT}
\newline
 This research was supported by the Asher Fund for
Space Research at the Technion.

\end{document}